\documentstyle[preprint,eqsecnum,aps]{revtex}
\scrollmode

\tightenlines

\begin{document}

\widetext

\draft

\baselineskip 20pt

\title{Off-shell effects in the electromagnetic production of strangeness}

\vspace{0.5cm}

\author{T. Mizutani$^1$, C. Fayard$^2$, G.-H. Lamot$^2$, and B. Saghai$^3$}

\address{1) Department of Physics,
Virginia Polytechnic Institute and State University\\
Blacksburg, VA 24061 USA}

\address{2) Institut de Physique Nucl\' eaire de Lyon, IN2P3-CNRS,
 Universit\'e
Claude Bernard,\\ F-69622 Villeurbanne Cedex, France} 

\address{3) Service de Physique Nucl\' eaire, CEA-Saclay,
F-91191 Gif-sur-Yvette, France}

\date{\today}

\maketitle

\begin{abstract}

Previous approaches to the photo- and electro-production of strangeness 
off the proton, based upon effective hadronic Lagrangians, are extended 
here to incorporate the so called {\it off-shell effects} inherent 
to the fermions with spin $\ge 3/2$.
A formalism for intermediate-state, spin 3/2, nucleonic and hyperonic 
resonances is presented and applied to the processes
$\gamma p~\rightarrow~K^{+}\Lambda$, 
for $E_{\gamma}^{lab} \leq$ 2.5 GeV,  
$ep~{\rightarrow}~e'K^+\Lambda$, as well as
the branching ratio for the crossed channel reaction
$K^-p~\rightarrow~\gamma\Lambda$, with 
stopped kaons. The sensitivity, from moderate to significant, of various 
observables to such effects are discussed. 
\end{abstract}
\narrowtext

\newpage

\section{Introduction}

The purpose of the present  work is to improve the recent 
Saclay-Lyon (SL) study~\cite{SL} on the strangeness
electromagnetic production from the proton. This latter investigation 
was based upon an 
effective hadronic Lagrangian in the lowest (tree) approximation, 
often called the isobar approximation.  
In a number of aspects one might safely say that SL is an improved 
version of its predecessors dealing with the same strangeness production
processes\footnote{
See Ref.\cite{SL} for a 
detailed account on this matter and extensive references to 
relevant papers.
}.
In particular, it has incorporated 
the {\it s-}channel nucleonic resonances with spin 3/2 and 5/2,  
expected to be important should the model keep adequate as energy
increases. In Ref.~\cite{Ren} such resonances were also 
considered.  
However, there the components of the amplitude growing undesirably 
with increasing channel energy were taken away by hand.  

As we will see later, these contributions arise from the non-resonant terms
associated to each considered resonance with spin~$>$~1/2.
In the SL study this was avoided 
by modifying the vertices and propagators in a manner adopted for spin 3/2
resonances in Refs.~\cite{Adel1,Adel2}: a straightforward extension to higher 
spins, while preserving the electromagnetic gauge invariance.  

This 
modification, however, has introduced an unwanted behavior for 
spin $>1/2$ hyperonic resonances exchanged in the {\it u-}channel: the 
corresponding propagators become singular in the physical region.  
Thus in the SL approach only spin 1/2 hyperons have been considered in the 
{\it u-}channel exchange. 
The phenomenological success of the SL model might imply that, within
the present state of the data, the main contributions from baryonic higher
spin resonances come mainly from the {\it s}-channel resonances (we will come 
back to this point in section IV).   
  
In the study of pion photoproduction,
the Rensselaer Polytechnic Institute (RPI) group~\cite{Ben89,dav91} has shown 
that of several different forms of the 
spin 3/2 propagator in the literature only one of them has a correct 
inverse. Also the authors pointed out that there are extra degrees of 
freedom associated with the interaction vertices involving a spin 3/2 
particle.  
By exploiting these facts, they successfully fitted the 
existing photo-pion data by the amplitudes generated from effective 
hadronic Lagrangians, and made predictions for some observables as 
well as the $E2/M1$ ratio for the $N\Delta \gamma$ vertex.  A similar 
strategy has been applied also by the 
RPI group~\cite{Ben95,Ben92} to the photo- and electro-production of 
the $\eta$ meson.

It seems quite natural then, as an extension of the 
Saclay-Lyon approach~\cite{SL}, as well as the 
works of the RPI group~\cite{Ben89,dav91,Ben95,Ben92}, to exploit this 
treatment for 
spin 3/2 particles in the study of the photo- and electro-production of 
the strangeness off the nucleon.
Yet, one needs to incorporate properly the {\it u-}channel exchanges in
the phenomenological approaches. The reasons for such an effort are mainly
two-fold: 
{\it (i)} a consistent treatment of the higher spin baryonic 
resonances
in both {\it s-} {\it and} {\it u-}channels, 
{\it (ii)} very likely, more sophisticated formalisms will be needed to
interpret the forthcoming data from new facilities, {\it e.g.}, 
the Thomas Jefferson National Accelerator Facility (JLAB),
the  Electron Stretcher Accelerator (ELSA),
the European Synchrotron Radiation Facility (ESRF), and the 8~GeV 
Synchrotron facility  (SPring-8) under construction in Japan.

In this paper, we work out the general expressions valid for the processes
with a kaon $K$ ($\equiv$~$K^+,K^\circ$) and an 
hyperon $Y$ ($\equiv \Lambda, \Sigma^\circ, \Sigma^+$) in the final state.
A selected set of $K\Lambda$ channel observables for the following processes 
are also reported: $\gamma p~\rightarrow~K^{+}\Lambda$ 
($E_{\gamma}^{lab} \leq$ 2.5 GeV),  
$ep~{\rightarrow}~e'K^+\Lambda$, and $K^-p~\rightarrow~\gamma\Lambda$.
Similar investigations with the $\Sigma$ hyperons in the final state, 
{\it i.e.} $K^+ \Sigma^\circ$ and $K^\circ \Sigma^+$ channels, 
are in progress and the results will be reported elsewhere.

In section II, the approach by the RPI group is extended to the
photo- and electroproduction of strangeness through {\it s-}channel
nucleonic resonances of spin 3/2. The off-shell parameters are 
introduced in the interaction Lagrangians, and the dependence 
on these parameters of the non-pole part of the invariant amplitudes  
is clarified. The approaches used previously where the off-shell
effects were ignored are placed in the present context. 
Section III is devoted to the treatment of spin 3/2 resonances 
in the {\it u-}channel.
The direct calculation proceeds along the same line as for the 
{\it s-}channel
resonance exchange. The substitution rule which emerges from the
direct calculation is worked out, leading to simple rules to
obtain the {\it u-}channel invariant amplitudes from the 
{\it s-}channel ones.
In section IV, we give our results and we discuss the dependence 
of the relevant observables on the off-shell parameters. 
The summary and conclusions are presented in the last section.

\newpage

\section{Spin 3/2 Resonances in the {\it s}-Channel}

In this section we extend the approach by Benmerrouche {\it 
et al.}~\cite{Ben89,dav91,Ben95}, devoted to the $\pi$ and $\eta$
photoproduction,  to obtain the amplitudes for the reactions 
$\gamma_{_{R,V}} p \to K Y$ ($K Y \equiv K^+ \Lambda , K^+ \Sigma^\circ ,  
K^\circ \Sigma^+$) for both real ($\gamma_{R}$) and virtual ($\gamma_{V}$)
photons, through an {\it s-}channel nucleonic  resonance of spin 3/2 
and positive parity.  
Once we obtain the amplitude, it is easy to establish its relation to 
the corresponding one obtained by Renard \& Renard~\cite{Ren} as well 
as to the 
one in SL~\cite{SL}.  Also one finds that the amplitudes due to  
nucleonic {\it s-}channel $3/2^-$ resonances may be trivially obtained by 
simple substitutions.  
Although some parts of this section should appear to be repetitive to 
those who are familiar with Ref.\cite{Ben89}, we give a 
comprehensive presentation of the matter for completeness, and present 
the explicit expressions of the invariant amplitudes for the photo-
and electro-production.

\subsection{Free Lagrangian}

We first define\footnote{
Throughout the present 
article we follow the conventions found in Bjorken and Drell\cite{BD}.
} 
the nucleon field as $N$ and a spin 3/2 (isospin 1/2) 
vector spin field (resonance) as $R^{\mu}$. Then the 
free Lagrangian for the spin 3/2 field reads
  
\begin{equation}
{\cal L}_{free} = \overline R^{\alpha} \Lambda_{\alpha \beta}R^{\beta}, 
\end{equation}

\noindent
where $\overline R^{\alpha}$ is the usual Dirac conjugate of $R^{\alpha}$, and
\begin{eqnarray}
\Lambda_{\alpha \beta} 
&=& -[(-i{\partial} \hskip -0.24 cm /  + M_R)g_{\alpha \beta} -iA
(\gamma_{\alpha} \partial_{\beta} +\gamma_{\beta} \partial_{\alpha})
-\frac{i}{2}(3A^2+2A+1)
\gamma_{\alpha} {\partial} \hskip -0.24 cm /  \gamma_{\beta} 
                                                             \nonumber \\
&& - M_R(3A^2+3A+1) \gamma_{\alpha} \gamma_{\beta}],
\end{eqnarray}

\noindent with $M_R$ the mass of the resonance, and $A(\ne -1/2)$ 
being a free parameter which preserves the invariance of the physical 
quantities constructed from the field under the point transformation

\begin{eqnarray}
R^{\mu} \to R^{\mu} + a\gamma^{\mu} \gamma^{\nu} R_{\nu},
\label{eq:point}\\
A \to A + (A-2a)/(1+4a),
\end{eqnarray}

\noindent $a \ne -1/4$, but otherwise arbitrary.  The free spin 3/2 
field satisfies the equation of motion and two constraints

\begin{eqnarray}
(i{\partial} \hskip -0.24 cm /  - M_R)R^{\mu} = 0,\\
\gamma_{\mu} R^{\mu} = 0,\\
\partial_{\mu} R^{\mu} = 0.
\end{eqnarray}

\noindent The above constraints ensure that $R$ has spin 3/2 with upper 
(positive energy) and lower (negative energy) components. 

The propagator associated with the $R$ field is obtained from the 
equation
\begin{equation}
\Lambda_{\alpha \beta} P^{\beta}_{\delta} = g_{\alpha \delta}.
\label{eq:prop}
\end{equation}

\noindent We may set $A= -1$ to find the simplest form for the propagator

\begin{equation}
P_{\mu \nu}(q) =
\frac{{q}\hskip -0.19 cm / + M_R}{3(q^2 - M_R^2)}\left[3g_{\mu \nu} 
-\gamma_{\mu} \gamma_{\nu} - \frac{2q_{\mu} q_{\nu}}{M_R^2} - 
\frac{q_{\nu} \gamma_{\mu} -q_{\mu} \gamma_{\nu}}{M_R}\right],
\label{eq:prop32}
\end{equation}

\noindent where $q$ is the four momentum of the resonance. It is important to 
note \cite{Ben89} that this propagator contains 
the spin 1/2 contribution, which is a consequence of the fact that the 
above $P^{\mu \nu}$ has the correct inverse. 

\subsection{Interaction Lagrangians}

Now we introduce the interactions for $\gamma_{_{R,V}} p \to K  Y$ 
through the {\it s}-channel spin 3/2 resonance discussed above.  Again 
following Ref. \cite{Ben89} with some modifications appropriate for the 
processes under consideration, the most general interaction Lagrangian 
which preserves the symmetry under the point interaction introduced in 
the previous Subsection reads

\begin{eqnarray}
{\cal L}_{K Y R} &=& \frac{g_{K Y R}}{M_K}
\left[\overline R^{\nu} \Theta_{\nu\mu}(Z) Y \partial^{\mu} K + 
\overline Y (\partial^{\mu} K^{\dagger})\Theta_{\mu\nu}(Z)R^
{\nu}\right], \label{eq:intlag1} \\
{\cal L}_{\gamma p R}^{(1)} &=& \frac{i e g_1}{2 M_p}
\left[\overline R^{\nu} \Theta_{\mu\lambda}(Y) \gamma_{\nu}\gamma^5 
N F^{\nu\lambda} +
\overline N \gamma^5\gamma_{\nu} \Theta_{\lambda\mu}(Y) R^{\mu} 
F^{\nu\lambda} \right], \label{eq:intlag2} \\
{\cal L}_{\gamma p R}^{(2)} &=& \frac{- e g_2}{4 M_p^2}   
\left[\overline R^{\mu} \Theta_{\mu\nu}(X) \gamma^5 
(\partial_{\lambda} N) F^{\nu\lambda} -
(\partial_{\lambda} \overline N) \gamma^5 \Theta_{\nu\mu}(X) R^{\mu}
F^{\nu\lambda} \right].
\label{eq:intlag3}
\end{eqnarray}

\noindent
In expression (\ref{eq:intlag1}), ${\cal L}_{K Y R}$ specifies the 
Lagrangian for the strong Kaon-Hyperon-Resonance ($K Y R$) vertex in 
which $K$ denotes the iso-doublet

\begin{eqnarray*}
K=\pmatrix{ K^+ \cr
            K^0 \cr }.
\end{eqnarray*}

\noindent
${\cal L}^{(1)}$ and ${\cal L}^{(2)}$ are for the $\gamma^5$ and 
{\it derivative} electromagnetic coupling terms, respectively. 
There $F^{\mu \nu}$ is the standard electromagnetic field
tensor, and $\Theta_{\mu \nu}$ is defined as (for our choice of $A= -1$ in 
the previous Subsection)

\begin{eqnarray}
\Theta_{\mu\nu}(V) = g_{\mu\nu} - (V+\frac{1}{2}) \gamma_{\mu} \gamma_{\nu}. 
\label{eq:teta}
\end{eqnarray}

\noindent It is important to stress that in the above Lagrangian, $V 
=X, Y, Z$ are arbitrary parameters which conserve the symmetry of the 
free Lagrangian under the 
point transformation [Eq.~(\ref{eq:point})], and are often called 
the {\it off-shell parameters}.  
As will become clear later, we shall 
exploit this extra freedom to make the kaon electromagnetic production 
amplitudes well tamed.  In what follows we shall rather use 
$\widetilde X \equiv X + \frac{1}{2}$, 
$\widetilde Y \equiv 2 Y + 1$, 
$\widetilde Z \equiv Z + \frac{1}{2}$.

The contribution to the $S$-matrix from the {\it s}-channel resonance pole 
through the $\gamma^5$ coupling reads

\begin{eqnarray}
S_2^{(s)} = \frac{i^2}{2!} \int d^4x_2 d^4x_1\ 
T\{ {\cal L}_{KY R}(x_2) {\cal L}^{(1)}_{\gamma p R}(x_1) \},
\end{eqnarray}

\noindent
where $T$ is the time-ordering operator.
Using the above Lagrangians, the matrix element 
for the $\gamma^5$ term is obtained as

\begin{eqnarray}
<Y K | T_s^{(1)} | \gamma p> &=& 
-iG_1
\overline U_{Y}({\hbox{\boldmath{$p$}\unboldmath}}_{Y})
i p_K^{\eta}\Theta_{\eta\mu}(Z)  P^{\mu\nu}(q) \nonumber \\
&&\times \Theta_{\nu\chi}(Y) \gamma_{\beta} \gamma^5
[ -i p_{\gamma}^{\beta} \epsilon^{\chi} + i \epsilon^{\beta} 
p_{\gamma}^{\chi}]
U_p({\hbox{\boldmath{$p$}\unboldmath}}_{p}),
\label{eq:gamma5}
\end{eqnarray}

\noindent
where we have introduced the coupling constant

\begin{equation}
G_1\equiv \frac{eg_1g_{K Y R}}{2M_p M_K},
\label{eq:G1}
\end{equation}
$\epsilon^{\chi}$ is the polarization vector of the photon, $q=p_{\gamma}+p_p 
= p_K+p_{Y}$ is the total momentum ($s=q^2$), and $P^{\mu\nu}(q)$ 
is the spin-3/2 propagator introduced in the last 
Subsection, Eq.(\ref{eq:prop32}).

Using expression (\ref{eq:teta}) for  $\Theta_{\mu\nu}$ to calculate
the terms on both sides of the propagator, we find

\begin{eqnarray}
<Y K | T_s^{(1)} | \gamma p> &=& 
-i G_1 
\overline U_{Y}({\hbox{\boldmath{$p$}\unboldmath}}_{Y}) 
[(p_K)_{\mu} - \widetilde Z {p} \hskip -0.16 cm / _K \gamma_{\mu}]
 P^{\mu\nu}(q) \nonumber \\
&&\times \left\{ 
[\epsilon_{\nu}{p} \hskip -0.16 cm / _{\gamma}
-(p_{\gamma})_{\nu}{\epsilon} \hskip -0.16 cm / ] -
\widetilde Y \gamma_{\nu} 
[ {\epsilon} \hskip -0.16 cm / {p} \hskip -0.16 cm / _{\gamma}
-\epsilon {\cdot} p_{\gamma}] 
\right\} \gamma^5 U_p({\hbox{\boldmath{$p$}\unboldmath}}_{p}).
\label{eq:dirac}
\end{eqnarray} 

\noindent
A similar calculation leads to the {\it derivative} coupling contribution
corresponding to ${\cal L}^{(2)}$

\begin{eqnarray}
<Y K | T_s^{(2)} | \gamma p> &=& 
-i G_2 
\overline U_{Y}({\hbox{\boldmath{$p$}\unboldmath}}_{Y}) 
[(p_K)_{\mu} - \widetilde Z {p} \hskip -0.16 cm / _K \gamma_{\mu}]
 P^{\mu\nu}(q) \nonumber \\ 
&&\times \left\{[\epsilon {\cdot} p_p(p_{\gamma})_{\nu} -  
p_{\gamma} {\cdot} p_p \,\epsilon_{\nu}] +
\widetilde X \gamma_{\nu} 
[ p_{\gamma} {\cdot} p_p{\epsilon} \hskip -0.16 cm /   -
                     \epsilon {\cdot} p_p{p} \hskip -0.16 cm / _{\gamma} ]
\right\} \gamma^5 U_p({\hbox{\boldmath{$p$}\unboldmath}}_{p}),   
\label{eq:pauli}
\end{eqnarray}

\noindent with
\begin{equation}
G_2 \equiv \frac{eg_2g_{KY R}}{4M_p^2 M_K}.
\label{eq:G2}
\end{equation}

\subsection{Vertex functions}

Adding the above two contributions given in Eqs. (\ref{eq:dirac}) 
and (\ref{eq:pauli}), 
we can write the scattering amplitude $M_{fi}$ corresponding to the 
{\it s-}channel exchange of an $S^P=3/2^+$ resonance as

\begin{equation}
M_{fi}^{(s)} =
\overline U_{Y}({\hbox{\boldmath{$p$}\unboldmath}}_{Y})
 {\cal{V}}^{\mu}(K Y R)\ P_{\mu \nu}(q)
 {\cal{V}}^{\nu}(R p \gamma)\ U_p({\hbox{\boldmath{$p$}\unboldmath}}_{p}),
\label{eq:mfi}
\end{equation}

\noindent where the $(K Y R)$ vertex reads

\begin{equation}
{\cal{V}}^{\mu}(K Y R) = -\frac{g_{K Y R}}{M_K}
[\ p_K^{\mu} - {\widetilde Z} {p} \hskip -0.16 cm / _K \gamma^{\mu} ] ,
\label{eq:vertk} 
\end{equation}

\noindent
and the $(R p \gamma)$ vertex is

\begin{eqnarray}
{\cal{V}}^{\nu}(R p \gamma) &=&  \left[ 
\frac{eg_1}{2M_p} {\left(
\epsilon^{\nu}{p} \hskip -0.16 cm / _{\gamma} 
- p_{\gamma}^{\nu}{\epsilon} \hskip -0.16 cm / 
        -{\widetilde Y} \gamma^{\nu}
    ({\epsilon} \hskip -0.16 cm / {p} \hskip -0.16 cm / _{\gamma} 
    - \epsilon {\cdot} p_{\gamma})
     \right)} \right. \cr  
&+&\left. \frac{eg_2}{4M_p^2} 
{\left( \epsilon {\cdot} p_p p_{\gamma}^{\nu} - p_\gamma {\cdot} p_p 
\epsilon^{\nu}
+{\widetilde X} \gamma^{\nu}
(p_\gamma {\cdot} p_p {\epsilon} \hskip -0.16 cm /  - \epsilon 
{\cdot} p_p {p} \hskip -0.16 cm / _{\gamma}) \right)} 
 \right] i\gamma^5.
 \label{eq:vert32} 
\end{eqnarray} 
Note that, for the general case of electroproduction, the above vertex 
must be multiplied by $F^{R}=F_2^p$, the second Dirac 
form factor of the proton.  
In the case of photoproduction, this factor reduces to 
unity, and in addition we have $\epsilon {\cdot} p_{\gamma}=0$.

\subsection{Invariant amplitudes}

The Lorentz invariant matrix element for electroproduction is written
as

\begin{equation}
M_{fi}^{(s)}  =
i\ \overline U_{Y} \left( \sum_{j=1}^6 {\cal A}_j {\cal M}_j 
                                      \right) U_p,
\label{eq:mfigen}
\end{equation}

\noindent
where $\overline U_{Y}$ and $U_p$ are the spinors of the hyperon 
and the proton, respectively, ${\cal A}_j$'s are Lorentz invariant 
scalar functions of
the Mandelstam variables, and ${\cal M}_j$'s are the six usual gauge 
invariant matrices for the electroproduction
 
\begin{eqnarray}
{\cal M}_1 & = & \gamma^5 \, 
({p} \hskip -0.16 cm / _{\gamma}\;{\epsilon} \hskip -0.16 cm /  -
\epsilon {\cdot} p_{\gamma}),
\cr
{\cal M}_2 &=& 
2\gamma^5(\epsilon {\cdot} p_p\;p_{\gamma} {\cdot} p_{Y} -
\epsilon {\cdot} p_{Y}\;p_{\gamma} {\cdot} p_p),
\cr
{\cal M}_3 &=& \gamma^5
({\epsilon} \hskip -0.16 cm / \;p_{\gamma} {\cdot} p_p -
{p} \hskip -0.16 cm / _{\gamma}\;\epsilon {\cdot} p_p),
\cr
{\cal M}_4 &=&  
\gamma^5({\epsilon} \hskip -0.16 cm / \;p_{\gamma} {\cdot} p_{Y} -
{p} \hskip -0.16 cm / _{\gamma}\;\epsilon {\cdot} p_{Y}),
\cr
{\cal M}_5 & = & \gamma^5 \, ( p^2_{\gamma}\; {\epsilon} \hskip -0.16 cm /  -
\epsilon {\cdot} p_{\gamma}\; {p} \hskip -0.16 cm / _{\gamma}), 
\cr
{\cal M}_6 & = & 
\gamma^5 \, ( p^2_{\gamma}\; \epsilon {\cdot} p_{Y} -
\epsilon {\cdot} p_{\gamma}\;  p_{\gamma} {\cdot} p_{Y}) -
\gamma^5 \, ( p^2_{\gamma}\; \epsilon {\cdot} p_p -
\epsilon {\cdot} p_{\gamma}\;  p_{\gamma} {\cdot} p_p) .
\label{eq:mj}
\end{eqnarray}

\noindent 
Due to the second term, 
$-\gamma^5 \, ( p^2_{\gamma}\; \epsilon {\cdot} p_p -
\epsilon {\cdot} p_{\gamma}\;  p_{\gamma} {\cdot} p_p)$, 
the choice of ${\cal M}_6$  is different from that used 
in Refs.~\cite{Adel1} and \cite{SL}. This results in a few modifications
in the expressions of the CGLN amplitudes for the electroproduction as given
in Appendix~\ref{app:cgln}. The advantage of this choice is 
its symmetric property under the 
exchange $p_p \leftrightarrow -p_{Y}$, thus leading to more 
transparent relations between the {\it s-} and {\it u-}channels amplitudes, as 
shown in the next section.  In the case of photoproduction, 
($p_{\gamma}^2=0$, $\epsilon {\cdot} p_{\gamma}=0$), only the first 
four invariant amplitudes in Eq.  (\ref{eq:mj}) are needed.

Using the above expressions for the propagator and vertices, 
application of the Dirac algebra leads to the invariant amplitudes 
${\cal A}_j$, which are expressed as sums of resonant or
pole ($P$) 
and non-pole ($NP$) contributions.
In the case of the photoproduction we find

\begin{equation}
{\cal A}_j = \sum_{i=1}^2 
               G_i \left[\frac{P_{ij}^P}{s-M_R^2} + P_{ij}^{NP}\right]
                                      \quad , \quad (j=1,\ldots,4), 
\label{eq:ajphoto}
\end{equation}    
 
\noindent
The expressions of the $P_{ij}^{P,NP}$ coefficients are given in
Appendix~\ref{app:invamp}, Eqs.~(\ref{eq:p1j}) to (\ref{eq:p2j}).

The electroproduction amplitudes can be written in a similar form

\begin{equation}
{\cal A}_j = \sum_{i=1}^2 
               G_i \left[\frac{E_{ij}^P}{s-M_R^2} + E_{ij}^{NP}\right]
                                      \quad , \quad (j=1,\ldots,6) . 
\label{eq:ajelec}
\end{equation} 

\noindent
For $j=1,...,4$, the $E_{ij}^{P,NP}$ coefficients are expressed in 
terms of the above defined photoproduction coefficients $P_{ij}^{P,NP}$ as

\begin{equation}
    E_{ij}^{P,NP} = P_{ij}^{P,NP} + p_{\gamma}^2 \, R_{ij}^{P,NP}
         \quad , \quad (i=1,2) \quad , \quad (j=1,\ldots,4).
\label{eq:decomp}                     
\end{equation}

\noindent
The extra terms $R_{ij}^{P,NP}$ are given in 
Appendix~\ref{app:invamp}, Eqs.~(\ref{eq:r1j}) and (\ref{eq:r2j}).
Note that this decomposition is not necessary for 
$j=5,6$. The corresponding coefficients $E^{P,NP}_{ij}$ are given in
Eqs.~(\ref{eq:e1-56}) and (\ref{eq:e2-56}) of Appendix~\ref{app:invamp}.

Note that in the calculation of the observables (Sec. IV), the 
following replacement is made in the denominator of the pole 
contribution in Eqs.~(\ref{eq:ajphoto}) and (\ref{eq:ajelec})

\begin{equation}
s -M_R^2 \to s - M_R^2 + iM_R\Gamma_R,
\end{equation}

\noindent where $\Gamma_R$ is the width of the resonance. 

It should be important to emphasize here that the {\it pole} contributions 
(see Appendix~\ref{app:invamp}) are completely {\it independent} of $V 
(=X, Y, Z)$, hence with no off-shell dependence.
 
So far we have discussed the case in which the parity of the 
{\it s}-channel resonance is positive. 
For a resonance with negative 
parity, we have only to make the following replacements: ${\cal 
{V}}^{\mu}(KY R) \to i\gamma^5 {\cal {V}}^{\mu}(KY R)$ in 
Eq.~(\ref{eq:vertk}), and $i\gamma^5 \to 1$ in Eq.~(\ref{eq:vert32}).  
In the corresponding $M_{fi}$ amplitude [Eq.~(\ref{eq:mfi})], 
$\gamma^5$ is now acting onto the left of the first vertex.
Using the anti-commutation property of $\gamma^5$ with $\gamma^{\mu}$, 
it is easy to move the $\gamma^5$ matrix in the same position as in 
the positive parity case, namely onto the right of the second vertex.  
By inspection, we immediately obtain the parity rule for the invariant 
(pole and non-pole) amplitudes

\begin{equation}
E_{ij}^{(-)}(M_R) = (-)^{i+1} E_{ij}^{(+)}(-M_R) \quad , \quad
(i=1,2) \quad , \quad (j=1,...,6).
\end{equation}

\subsection{Formalisms without off-shell effects}

        \subsubsection{Renard and Renard approach}

The expressions used in 
Ref. \cite{Ren} for the propagator is 
the same as Eq.~(\ref{eq:prop32}), but in the interaction Lagrangian 
Eqs.~(\ref{eq:intlag1})-(\ref{eq:intlag3}) 
$\Theta_{\mu \lambda}(V),\ (V =X, Y, Z)$ was set 
equal to $g_{\mu \lambda}$.
In other words, the authors put $V \equiv -\frac{1}{2}$ in 
(\ref{eq:teta}) (or $\widetilde V \equiv 0$), thus no off-shell effect 
associated with the spin 3/2 particles was considered.  It is 
therefore clear from Eqs.~(\ref{eq:vertk}) and (\ref{eq:vert32}) that 
the corresponding amplitude simplifies considerably.  
However, some of the {\it non-pole} 
contributions $P_{ij}^{NP}$ grow linearly in the {\it s-}variable
(see Appendix~\ref{app:invamp}), causing 
an undesirable increase, for example, in the production cross section.  
For this reason {\it all} the resulting {\it non-pole} contributions were 
artificially thrown away in Ref. \cite{Ren}.

           \subsubsection{Adelseck et al. approach}

To avoid the difficulties encountered in the Renard \& Renard model 
\cite{Ren}, Adelseck {\it et al.}~\cite{Adel1} 
have suggested and applied the following 
prescriptions (used also in Ref.~\cite{SL}).  
The propagator is written from Eq.  (\ref{eq:prop32}), 
with the mass of the resonance $M_R$ replaced by the total invariant 
energy $\sqrt s$, except in the denominator where the width of
the resonance is introduced

\begin{equation}
P_{\mu\nu}^A =
 \frac{{q \hskip -0.19 cm / }+\sqrt{s}}
{3(s-M_R^2+iM_R\Gamma_R)}
\left[3g_{\mu\nu}-\gamma_{\mu}\gamma_{\nu}-\frac{2}{s}q_{\mu}q_{\nu}
-\frac{1}{\sqrt{s}}(\gamma_{\mu}q_{\nu}-\gamma_{\nu}q_{\mu})\right].
\label{eq:prop32Adel}
\end{equation}

\noindent This modification provides an extra damping of the 
amplitude with increasing channel energy.  So together with the 
corresponding modification in the vertices 
discussed below, an unwanted growth in 
the production cross section due to the non-pole contribution could be 
reduced in the absence of the off-shell freedom (in terms of 
$X, Y, Z$).

In the photoproduction case, the $K Y R$ vertex is 

\begin{equation}
{\cal{V}}^{\mu}(K Y R) = \frac{\widetilde g_{KY R}}{M_R}
\ p_{Y}^{\mu} ,
\label{eq:vkln32Adel}
\end{equation}

\noindent
and the $R p \gamma$ vertex factor 
for a positive parity resonance is written as

\begin{equation}
{\cal{V}}^{\nu}(R p \gamma) = i\ \left[
g_a \left(\epsilon^{\nu}
- \frac{p_{\gamma}^{\nu}{\epsilon \hskip -0.16 cm / }}
{\sqrt{s}+M_p}\right)
+g_b \frac{1}{(\sqrt{s}+M_p)^2}(\epsilon 
{\cdot}p_p p_{\gamma}^{\nu} -p_\gamma{\cdot}p_p \epsilon^{\nu})
\right] \gamma^5 .
\label{eq:vn32Adel}
\end{equation} 

As stated by Adelseck {\it et al.}, these prescriptions were used to 
ensure gauge invariance of the scattering amplitude.  In fact, 
expressions (\ref{eq:vkln32Adel}) and (\ref{eq:vn32Adel}) may be reached 
from Eqs.~(\ref{eq:vertk}) and (\ref{eq:vert32}) as demonstrated in 
Appendix~\ref{app:nooff}, 
where the coupling constants $g_a$, $g_b$, and $\widetilde g_{KY R}$ 
are defined in terms of $g_1$, $g_2$, and $g_{K Y R}$, 
respectively.  Particularly, the photon coupling vertex in this choice 
contains damping factors in the {\it s-}variable.  

 However, regarding the spin 3/2 propagator (\ref{eq:prop32Adel}), 
 when the same form is used for a {\it u-}channel resonance exchange, 
 namely the {\it s-}variable replaced by the {\it u-}variable, the 
latter may 
 vanish at certain kinematical situations, leading to an unphysical 
 behavior.  Note also that as pointed out in \cite{Ben89}, 
such propagators do not have inverses, and corresponding wave equations 
for the the spin-3/2 field can not be defined.
Thus this approach is not 
appropriate for a consistent simultaneous description of the {\it s-} and 
 {\it u-}channels.

  
\newpage

\section{Spin $3/2$ resonances in the {\it u-}channel}

       \subsection{Direct calculation}

In this section we show some basic details on how the 
lowest order {\it u-}channel exchange of a $\Lambda^*(3/2^+)$ resonance 
contributes to the amplitude for $K^+$ production on the proton.  The 
exchange of a $\Lambda^*(3/2^+)$ resonance in the {\it u-}channel is 
treated along the same lines as in section II for the 
{\it s-}channel resonances exchange.  The part of the $S$-matrix 
corresponding to the {\bf $\gamma^5$} photon coupling is

\begin{eqnarray}
S_2^{(u)} = \frac{i^2}{2!} \int d^4x_2 d^4x_1 \
T\{ {\cal L}^{(1)}_{\gamma Y R}(x_2) 
    {\cal L}_{K p R}(x_1) \},
\end{eqnarray}

\noindent
with $R \equiv \Lambda^*(3/2^+)$. The resulting matrix element takes the form

\begin{eqnarray}
<Y K | T_u^{(1)} | \gamma p> = 
\frac{-i e g'_1 g_{K p R}}{2 M_{Y} M_K} 
\overline U_{Y}({\hbox{\boldmath{$p$}\unboldmath}}_{Y})  
\gamma_5\gamma_{\beta}
&&[ -i p_{\gamma}^{\beta} \epsilon^{\lambda} + i \epsilon^{\beta} p_{\gamma}^
{\lambda}]
\Theta_{\lambda\nu}(Y) \nonumber\\
&&\times P^{\nu\mu}(-q')
\Theta_{\mu\chi}(Z) i p_K^{\chi}
U_p({\hbox{\boldmath{$p$}\unboldmath}}_{p}).
\end{eqnarray}

\noindent The momentum transfer is 
$q'\equiv p_{\gamma}-p_{Y} = 
p_K-p_p$, with $q'\,^2=u$.  Note that with a correct kinematical 
consideration it is easy to see that the propagator depends on $-q'$ 
(not $q'$ !).

Using Eq.~(\ref{eq:teta}) just as before, one finds

\begin{eqnarray}
<Y K | T_u^{(1)} | \gamma p> = 
\frac{-ie g'_1 g_{K p R}}{2 M_{Y} M_K}
\overline U_{Y}({\hbox{\boldmath{$p$}\unboldmath}}_{Y})\gamma_5 && \left\{
[\epsilon_{\nu}{p} \hskip -0.16 cm / _{\gamma}
-(p_{\gamma})_{\nu}{\epsilon} \hskip -0.16 cm / ] -
\widetilde Y 
[ {p} \hskip -0.16 cm / _{\gamma}{\epsilon} \hskip -0.16 cm / 
-\epsilon {\cdot} p_{\gamma}]
\gamma_{\nu} \right \} \nonumber \\
&& \times P^{\nu\mu}(-q') [(p_K)_{\mu} - \widetilde Z \gamma_{\mu} 
{p} \hskip -0.16 cm / _K ]
U_p({\hbox{\boldmath{$p$}\unboldmath}}_{p}).
\label{eq:u-dirac}
\end{eqnarray}

\noindent The {\it derivative} coupling term is calculated along the 
same lines, leading to

\begin{eqnarray}
<Y K | T_u^{(2)} | \gamma p> = 
\frac{-i e g'_2 g_{K p R}}{4 M_{Y}^2 M_K}
\overline U_{Y}({\hbox{\boldmath{$p$}\unboldmath}}_{Y})\gamma_5 &&  \left\{
[\epsilon {\cdot} p_{Y}(p_{\gamma})_{\nu} -
p_{\gamma} {\cdot} p_{Y} \,\epsilon_{\nu}] + 
\widetilde X [ p_{\gamma} {\cdot} p_{Y}{\epsilon} \hskip -0.16 cm /   -
           \epsilon {\cdot} p_{Y} {p} \hskip -0.16 cm / _{\gamma} ]\gamma_{\nu}
\right\}\nonumber \\           
&& \times P^{\nu\mu}(-q') 
[(p_K)_{\mu} - \widetilde Z \gamma_{\mu}{p} \hskip -0.16 cm / _K ] 
U_p({\hbox{\boldmath{$p$}\unboldmath}}_{p}).
\label{eq:u-pauli}
\end{eqnarray}             

\noindent
In the above expressions, $g'_1$ and $g'_2$ are the two $\gamma Y R$
coupling constants, which are similar to $g_1$ and $g_2$ as in
Eqs.~(\ref{eq:intlag2}) and (\ref{eq:intlag3}). 
Note the similarity of the last two expressions with the 
corresponding ones for an {\it s-}channel resonance exchange, 
Eqs.~(\ref{eq:dirac}) and (\ref{eq:pauli}). Adding the two above 
contributions, the scattering matrix in the {\it u-}channel 
exchange reads

\begin{equation}
M_{fi}^{(u)} =
\overline U_{Y}({\hbox{\boldmath{$p$}\unboldmath}}_{Y})
\ {\cal{V}}^{\nu}(R Y \gamma)\ P_{\nu \mu}(-q')
\ {\cal {V}}^{\mu}(K p R )\ U_p({\hbox{\boldmath{$p$}\unboldmath}}_{p}).
\label{eq:u-mfi}
\end{equation}

\noindent
The two vertices are

\begin{equation}
\ {\cal{V}}^{\mu}(K p R) = -\frac{g_{K p R}}{M_K}
[\ p_K^{\mu} - {\widetilde Z} \gamma^{\mu} {p} \hskip -0.16 cm / _K  ],
\label{eq:u-vertk} 
\end{equation}

\begin{eqnarray}
{\cal {V}}^{\nu}(R Y \gamma) &=& i \gamma_5 \left[ 
\frac{e g'_1}{2M_{Y}} {\left(
\epsilon^{\nu}{p} \hskip -0.16 cm / _{\gamma} 
- p_{\gamma}^{\nu}{\epsilon} \hskip -0.16 cm / 
        -{\widetilde Y}
    ({p} \hskip -0.16 cm / _{\gamma}{\epsilon} \hskip -0.16 cm /  
    - \epsilon {\cdot} p_{\gamma}) 
                         \gamma^{\nu}\right)} 
\right. \cr  
&+&\left. \frac{e g'_2}{4M^2_{Y}} 
{\left( \epsilon {\cdot} p_{Y} p_{\gamma}^{\nu} - 
p_\gamma {\cdot} 
p_{Y} \epsilon^{\nu}
+{\widetilde X} 
(p_\gamma {\cdot} p_{Y} {\epsilon} \hskip -0.16 cm /  - \epsilon {\cdot} 
p_{Y} 
{p} \hskip -0.16 cm / _{\gamma})
\gamma^{\nu}\right)} 
\right]. 
\label{eq:u-vert32} 
\end{eqnarray} 

\noindent The propagator reads

\begin{equation}
P_{\nu\mu}(-q') =
 \frac{-{q'}\hskip -0.29 cm /+M_R}{3(u-M_R^2)}
\left[3g_{\nu\mu}-\gamma_{\nu}\gamma_{\mu}-
                              \frac{2}{M_R^2}q'_{\nu}q'_{\mu}
+\frac{1}{M_R}(\gamma_{\nu}q'_{\mu}-\gamma_{\mu}q'_{\nu})\right].
\label{eq:u-prop32}
\end{equation}

Using the above expressions for the vertices and propagator, the 
decomposition of $M_{fi}^{(u)}$ in terms of the gauge invariant 
matrices defined in Eq.~(\ref{eq:mj}) can be done along the same 
lines as in section II.D. However, comparing the {\it s-} and 
{\it u-}channels 
vertices and propagators, it is easy to get out the rules regarding 
how to obtain the expressions for the {\it u-}channel exchange 
from those for the {\it s-}channel, namely:
{\it 1)} exchange $p_p \leftrightarrow -p_{Y}$ (including $M_N 
\to M_{Y}$),
{\it 2)}~express the products of $\gamma$ matrices in a reversed order,
{\it 3)} change $s \rightarrow u$, $g_2 \rightarrow -g_2$,
$M_R \rightarrow -M_R$, and
{\it 4)} exchange the two vertices and put the appropriate
coupling constants. In fact, these rules result from a substitution rule
which is simpler to use, since it allows us to formally derive the invariant 
amplitudes for the {\it u-}channel exchange directly from the 
corresponding {\it s-}channel exchange amplitudes. The derivation of
the substitution rule and its application to obtain the invariant amplitudes
are given in the next two Subsections.  

     \subsection{Substitution rule}

We now calculate the {\it u-}channel results by substitution rule applied 
to the {\it s-}channel expressions.  Namely, we introduce 
in Eq.~(\ref{eq:mfi}) the appropriate coupling constants, 
together with the following replacements:
$s \to u$, $M_N \to M_{Y}$,
$p_p \leftrightarrow -p_{Y}$, 
$U_p({\hbox{\boldmath{$p$}\unboldmath}}_{p}) 
   \rightarrow V_{Y}(-{\hbox{\boldmath{$p$}\unboldmath}}_{Y})$,
$U_{Y}({\hbox{\boldmath{$p$}\unboldmath}}_{Y}) 
   \rightarrow V_p(-{\hbox{\boldmath{$p$}\unboldmath}}_{p})$,
with $V_p$, $V_{Y}$ being the negative energy spinors.
The resulting scattering matrix is 

\begin{equation}
M_{fi}^{(u)} =
\overline V_p(-{\hbox{\boldmath{$p$}\unboldmath}}_{p})
\ {\cal {V}}^{\mu}(K p R) \ P_{\mu \nu}(q')
\ {\cal {V}}^{\nu}(R Y \gamma) 
\ V_{Y}(-{\hbox{\boldmath{$p$}\unboldmath}}_{Y}),
\label{eq:mfi-u}
\end{equation}

\noindent
and the expressions of the vertices and propagator are
[cf. Eqs.~(\ref{eq:vertk}), (\ref{eq:vert32}), and (\ref{eq:prop32})]:

\begin{equation}
{\cal {V}}^{\mu}(K^ R) = -\frac{g_{K p R}}{M_K}
[\ p_K^{\mu} - {\widetilde Z} {p} \hskip -0.16 cm / _K \gamma^{\mu} ],
\label{eq:vertk-u} 
\end{equation}

\begin{eqnarray}
{\cal {V}}^{\nu}(R Y \gamma) &=& i \left[ 
\frac{e g'_1}{2M_{Y}} {\left(
\epsilon^{\nu}{p} \hskip -0.16 cm / _{\gamma} 
- p_{\gamma}^{\nu}{\epsilon} \hskip -0.16 cm / 
        -{\widetilde Y} \gamma^{\nu}
    ({\epsilon} \hskip -0.16 cm / {p} \hskip -0.16 cm / _{\gamma} - 
                              \epsilon {\cdot} p_{\gamma}) \right)} 
                                                          \right. \cr  
&-&\left. \frac{e g'_2}{4M^2_{Y}} 
{\left( \epsilon {\cdot} p_{Y} p_{\gamma}^{\nu} - 
                         p_\gamma {\cdot} p_{Y} \epsilon^{\nu}
+{\widetilde X} \gamma^{\nu}
(p_\gamma {\cdot} p_{Y} {\epsilon} \hskip -0.16 cm /  - 
             \epsilon {\cdot} p_{Y} {p} \hskip -0.16 cm / _{\gamma}) \right)} 
\right] \gamma_5,
\label{eq:vert32-u} 
\end{eqnarray}

\begin{equation}
P_{\mu\nu}(q') =
 \frac{{q'}\hskip -0.29 cm /+M_R}{3(u-M_R^2)}
\left[3g_{\mu\nu}-\gamma_{\mu}\gamma_{\nu}-
                              \frac{2}{M_R^2}q'_{\mu}q'_{\nu}
-\frac{1}{M_R}(\gamma_{\mu}q'_{\nu}-\gamma_{\nu}q'_{\mu})\right],
\label{eq:prop32-u}
\end{equation}

\noindent
with $q' = p_{\gamma} - p_{Y} = p_K - p_p$, as
before.

Using the relation between the $V$ and $U$ spinors:
$ V(-{\hbox{\boldmath{$p$}\unboldmath}}\,) = 
                      C \overline U^T({\hbox{\boldmath{$p$}\unboldmath}}\,)$,
with $C=\gamma_0\gamma_2$ the charge conjugation operator, and
$\overline U=U^{\dagger} \gamma_0$ the Dirac adjoint of $U$,
Eq.~(\ref{eq:mfi-u}) is written as:

\begin{equation}
M_{fi}^{(u)} = - U^T_p({\hbox{\boldmath{$p$}\unboldmath}}_{p})\ C^{-1}
\ {\cal {V}}^{\mu}(K p R) \ P_{\mu \nu}(q')
\ {\cal {V}}^{\nu}(R Y \gamma)\ 
C \ \overline U^T_{Y}({\hbox{\boldmath{$p$}\unboldmath}}_{Y}),
\label{eq:mfi-u2}
\end{equation}

By appropriately inserting $I=C^{-1}C$,  the above equation can be 
transformed into

\begin{equation}
M_{fi}^{(u)} = - \overline U_{Y}({\hbox{\boldmath{$p$}\unboldmath}}_{Y}) \ 
         \ [{\cal {V}}^{\nu}(R Y \gamma)^T]^C
         \ [P_{\mu \nu}(q')^T]^C
         \ [{\cal {V}}^{\mu}(K p R)^T]^C  
                 \ U_p({\hbox{\boldmath{$p$}\unboldmath}}_{p}),
\label{eq:mfi-u3}
\end{equation}

\noindent
where we have defined the $C$-transform of $X^T$ as

$$ [X^T]^C = C^{-1} X^T C. $$

\noindent Now, we exploit the properties of the charge conjugation 
matrix $C$ to calculate the $C$-transforms of the vertices and 
propagator. For example, the $C$-transform of the $(K p R)$ 
vertex Eq.(\ref{eq:vertk-u}) is

\begin{equation} 
[{\cal {V}}^{\mu}(K p R)^T]^C = -\frac{g_{K p R}}{M_K}  
  C^{-1}[\ p_K^{\mu} 
  - {\widetilde Z} {p} \hskip -0.16 cm / _K \gamma^{\mu} ]^T C.
\label{eq:c-vertk-u}
\end{equation}

\noindent
From the properties of $C$, we obtain
\begin{equation}
 C^{-1} ({p} \hskip -0.16 cm / _K \gamma^{\mu})^T C = 
 C^{-1} \gamma^{\mu\>T} (p_K)_{\nu} \gamma^{\nu\>T} C =
        \gamma^{\mu} (p_K)_{\nu} \gamma^{\nu} =
        \gamma^{\mu} {p} \hskip -0.16 cm / _K, 
\end{equation}

\noindent
and Eq.~(\ref{eq:c-vertk-u}) becomes

\begin{equation} 
[{\cal {V}}^{\mu}(K p R)^T]^C = -\frac{g_{K p R}}{M_K}  
  [\ p_K^{\mu} - {\widetilde Z} \gamma^{\mu} {p} \hskip -0.16 cm / _K ],
\label{eq:c-vertk-u2}
\end{equation}

A similar calculation leads to the $R Y \gamma$ vertex

\begin{eqnarray}
[{\cal {V}}^{\nu}(R Y \gamma)^T]^C &=& i \gamma_5 \left[ 
- \frac{e g'_1}{2M_{Y}}{\left(
\epsilon^{\nu}{p} \hskip -0.16 cm / _{\gamma} 
- p_{\gamma}^{\nu}{\epsilon} \hskip -0.16 cm / 
        -{\widetilde Y} 
    ({p} \hskip -0.16 cm / _{\gamma} {\epsilon} \hskip -0.16 cm / - 
                    \epsilon {\cdot} p_{\gamma})\gamma^{\nu} \right)} 
                                                          \right. \cr  
&-&\left. \frac{e g'_2}{4M^2_{Y}} 
{\left( \epsilon {\cdot} p_{Y} p_{\gamma}^{\nu} - 
                         p_\gamma {\cdot} p_{Y} \epsilon^{\nu}
+{\widetilde X} 
(p_\gamma {\cdot} p_{Y} {\epsilon} \hskip -0.16 cm /  - 
  \epsilon {\cdot} p_{Y} {p} \hskip -0.16 cm / _{\gamma})\gamma^{\nu} \right)} 
\right],
\label{eq:c-vert32-u} 
\end{eqnarray} 

\noindent
and the propagator takes the form

\begin{equation}
[P_{\mu\nu}(q')^T]^C =
 \frac{-{q'}\hskip -0.29 cm /+M_R}{3(u-M_R^2)}
\left[3g_{\nu\mu}-\gamma_{\nu}\gamma_{\mu}-
                              \frac{2}{M_R^2}q'_{\nu}q'_{\mu}
+\frac{1}{M_R}(\gamma_{\nu}q'_{\mu}-\gamma_{\mu}q'_{\nu})\right].
\label{eq:c-prop32-u}
\end{equation}

\noindent
Comparing with Eq.~(\ref{eq:prop32-u}) leads to

\begin{equation}
[P_{\mu\nu}(q')^T]^C = P_{\nu\mu}(-q').
\label{eq:cc-prop32-u}
\end{equation}

\noindent
Combining Eqs.~(\ref{eq:mfi-u3}), (\ref{eq:c-vertk-u2}),
(\ref{eq:c-vert32-u}) and (\ref{eq:cc-prop32-u}) leads to the
same result as the direct calculation 
Eqs.~(\ref{eq:u-mfi})-(\ref{eq:u-prop32}).

     \subsection{Invariant amplitudes}

In this subsection, we apply the substitution rule to obtain the invariant 
amplitudes (hereafter denoted as ${\cal A}'_j$) for the {\it u-}channel 
exchange directly from the corresponding {\it s-}channel exchange 
amplitudes ${\cal A}_j$.

Let us write Eq.~(\ref{eq:mfigen}) with specifying the relevant
variables

\begin{equation}
M_{fi}^{(s)}  =
i\ \overline U_{Y}({\hbox{\boldmath{$p$}\unboldmath}}_{Y})
   \left( \sum_{j=1}^6 
        {\cal A}_j(s,t,u) {\cal M}_j(p_p,p_{Y}) 
                           \right) U_p({\hbox{\boldmath{$p$}\unboldmath}}_{p}).
\label{eq:mfi-s}
\end{equation}

\noindent where ${\cal M}_j$ are the six gauge invariant amplitudes 
Eq.(\ref{eq:mj}), and  ${\cal A}_j$ have been  
made explicit in section II.D.  We now apply the substitution rule 
(see the previous subsection) to Eq.~(\ref{eq:mfi-s}) in order to 
obtain the scattering matrix in the {\it u-}channel

\begin{equation}
M_{fi}^{(u)}  =
i\ \overline V_p(-{\hbox{\boldmath{$p$}\unboldmath}}_{p})
   \left( \sum_{j=1}^6 
        {\cal A}_j(u,t,s) {\cal M}_j(-p_{Y},-p_p) 
                       \right)\ V_{Y}(-{\hbox{\boldmath{$p$}\unboldmath}}_{Y}).
\label{eq:mfi-u4}
\end{equation}

\noindent
From Eq.~(\ref{eq:mj}), it is clear that

\begin{eqnarray}
&&{\cal M}_{1,5,6}(-p_{Y},-p_p)={\cal M}_{1,5,6}(p_p,p_{Y}), 
\nonumber\\
&&{\cal M}_2(-p_{Y},-p_p) = -{\cal M}_2(p_p,p_{Y})
, \quad
{\cal M}_{3,4}(-p_{Y},-p_p)=-{\cal M}_{4,3}(p_p,p_{Y}).
\end{eqnarray}   

We proceed as in the last subsection, and transform 
Eq.~(\ref{eq:mfi-u4}) into

\begin{equation}
M_{fi}^{(u)}  = - i\ \overline U_{Y}({\hbox{\boldmath{$p$}\unboldmath}}_{Y})
 \left( \sum_{j=1}^6 
{\cal A}_j(u,t,s) \ C^{-1} {\cal M}^T_j(-p_{Y},-p_p)\ C \right) 
                    \ U_p({\hbox{\boldmath{$p$}\unboldmath}}_{p}),
\label{eq:mfi-u5}                    
\end{equation}

\noindent
Then, we calculate  the $C$-transforms
of the ${\cal M}^T_j(-p_{Y},-p_p)$ matrices, which can easily
be expressed in terms of the original 
${\cal M}_j(p_p,p_{Y})$ matrices as

\begin{eqnarray}
C^{-1} {\cal M}^T_{1,2}(-p_{Y},-p_p)\ C &=& 
     - {\cal M}_{1,2}(p_p,p_{Y}), \cr
C^{-1} {\cal M}^T_{5,6}(-p_{Y},-p_p)\ C &=& 
       {\cal M}_{5,6}(p_p,p_{Y}), \cr
C^{-1} {\cal M}^T_{3,4}(-p_{Y},-p_p)\ C &=&
       {\cal M}_{4,3}(p_p,p_{Y}).
\end{eqnarray} 

Substituting these relations into Eq.~(\ref{eq:mfi-u5}), the scattering
matrix in the {\it u-}channel can be written as

\begin{equation}
M_{fi}^{(u)}  =
i\ \overline U_{Y}({\hbox{\boldmath{$p$}\unboldmath}}_{Y})
   \left( \sum_{j=1}^6 
        {\cal A}'_j(s,t,u) {\cal M}_j(p_p,p_{Y}) 
                         \right) U_p({\hbox{\boldmath{$p$}\unboldmath}}_{p}),
\end{equation}

\noindent
where the invariant amplitudes ${\cal A}'_j$ are related to 
the ${\cal A}_j$ amplitudes in the {\it s-}channel as follows

\begin{equation}
{\cal A}'_{1,2}(s,t,u) =  {\cal A}_{1,2}(u,t,s) , \quad
{\cal A}'_{3,4}(s,t,u) =  {\cal A}_{4,3}(u,t,s) , \quad
{\cal A}'_{5,6}(s,t,u) = -{\cal A}_{5,6}(u,t,s).
\end{equation}

\noindent It is then quite easy to obtain  these invariant
amplitudes in a form similar to Eq.~(\ref{eq:ajelec}), which we will
not present in this article. 

  
\newpage
\section{Results and Discussion}

In this section we {\it illustrate} the sensitivity of different 
$K \Lambda$ channels observables
to the off-shell effects.
We need, hence, a reliable dynamical model, with respect to the 
existing data, as starting point. In the following,
we present first how a rather simple model was obtained and then, 
within the dynamical ingredients required by 
the available data, we report on the importance of
the off-shell effects according to the observables and/or the phase space
regions investigated. 


\subsection{Reaction mechanism}
To build a simple model with a reasonably realistic dynamical content,
we take advantage of the SL model \cite{SL} which has emerged from a 
comprehensive phenomenological study. 

The underlying dynamics in the SL model is, besides extended Born terms,
resonances exchanges (Table~\ref{tab:ABC}) in the following channels:

\begin{itemize}
\item{\it s-channel:} {$N1$(1/2), $N7$(3/2), $N8$(5/2) ; 
where the spin of each nucleonic
resonance is given in parenthesis.}
\item{\it u-channel: }{$L1$, $L3$, $L5$, $S1$ ; all spin 1/2 
hyperonic resonances.}
\item{\it t-channel:} {$K^*$, $K1$ ; both of them have also been
included in the present work and we will not discuss them any further.}
\end{itemize}
 
In the {\it s-}channel, the most relevant resonance, in the frame of
the present work, is the spin 3/2 resonance $N7$. The $N1$ resonance,
$P_{11}$(1440),
was found to have a coupling compatible with zero (see Table~\ref{tab:ABC} 
and Ref.~\cite{SL}).
Moreover, a recent 
model-independent nodal structure analysis~\cite{ST97} concludes that the 
present data do not require
contributions from the $P_{11}$ resonances. Concerning another nucleonic
resonance in the SL model, the $N8~(5/2)$, it was  
shown~\cite{SL} that its contribution to the considered underlying dynamics 
{\it is not crucial} (see Table XI in Ref.~\cite{SL}). 

For these reasons, {\it we removed
the $N1$ and $N8$ resonances in searching for a simple model to study the
role of off-shell effects}. The parameters of this model, hereafter 
called model A, 
have been obtained by re-fitting the data. 
Note that the formalism used in this re-fitting is still within the 
context of Adelseck {\it et al.}'s treatment for the spin 3/2 resonance $N7$. 
Model A is the basis
of our numerical results reported in the next subsection.

The first step was thus, using model A, to fit the same data base as used 
to obtain the SL model (all available 312 data points for photo-, 
electro-production, as well as for the $K^-p$ radiative capture
process). The coupling constants and the reduced $\chi^2$ are given
in Table~\ref{tab:ABC}. Although the resultant $\chi^2$ for model A 
(1.84) is slightly larger than that for the SL model (1.73), it is still
acceptable. 
Anticipating the presentation of the observables in the next
subsection, the fit of the data with model A appears at a comparable level 
of quality as with the SL model, 
see the dotted and dash-dotted curves in Figs.~\ref{fig:ds}-\ref{fig:pol}, 
and Fig.~\ref{fig:ul}. 
Hence, these results justify the use of model A as a starting point to
investigate the sensitivity of the observables to the off-shell effects.

Given that model A contains only one spin 3/2 baryonic resonance,
we have also investigated
possible contributions from other known spin 3/2 nucleonic 
resonances, namely\footnote{We use the notation [$(\ell)J^{\pi}$].}, 
[$N(1520)$[$(2){3\over 2}^-$] ($N2$) or
[$N(1700)$[$(2){3\over 2}^-$] ($N5$).
We performed minimizations for all possible configurations including one
to three of the spin 3/2 resonances $N2$, $N5$, and $N7$.
In these configurations, whenever at least one of the two resonances
$N2$, and $N5$ was retained, 
the corresponding $\chi^{2,}$s were found significantly larger than that for 
model A, implying that the existing 
data base does not require contributions from these resonances.
Through the numerical investigations mentioned above, we have re-confirmed 
that model A is
indeed a {\it reasonable starting model} for the present study. 

Then we 
adopted the correct propagator [Eq.~(\ref{eq:prop32})] and introduced the 
off-shell treatments to the $N7$ resonance, and fitted again the
data to obtained model B (Table~\ref{tab:ABC}).
Finally, for the sake of completeness 
we included the {\it u-}channel spin 3/2 hyperonic resonance 
[${\Lambda}(1890)$[$(0){3\over 2}^+$] ($L8$) with the 
off-shell effect,
and once again fitted the data (model~C in Table~\ref{tab:ABC}). 
The choice of this resonance, as in the case of nucleonic resonances 
mentioned above,
comes from the fact that the inclusion of any other spin 3/2 hyperonic 
resonance, 
[${\Lambda}(1520)$[$(0){3\over 2}^-$] ($L6$) or 
[${\Lambda}(1690)$[$(0){3\over 2}^-$] ($L7$), deteriorates the reduced 
$\chi^2$ significantly.

Here we would like to point out 
that by adding any spin 3/2 baryonic resonance we introduce
five additional free parameters, namely two coupling constants 
($G_1$ and $G_2$)  and three off-shell parameters. 
The fact that the $\chi^2$ associated with model C comes out larger 
than that for the model B, indicates that the dynamical 
content of the phenomenological approach discussed here is 
reliable enough, since {\it additional free parameters} due 
to apparently unrelevant resonances {\it do not improve} 
the $\chi^2$ (reduced or per point).

Model D in Table~\ref{tab:ABC}, with two of the free parameters fixed, 
will be discussed in Subsec.~C.


\subsection{Observables}
In this subsection, we compare with the data the results of the four  
dynamical models (SL, A, B, and C) summarized in Table~\ref{tab:ABC}. 
Here we will adhere closely to the observables reported for the SL 
model~\cite{SL}, 
where a comprehensive discussion on other available phenomenological 
results~\cite{ade90,wil92} is also presented.
%
%
\subsubsection {Reaction ${\gamma}~+~p~{\rightarrow}~K^+~+~{\Lambda}$}

In Fig.~\ref{fig:ds}, angular distributions and excitation functions for 
the differential cross-section are shown.
All the models reproduce the data almost equally well.
However, the excitation functions at 
$\theta_{K}^{cm}$ =  90$^{\circ}$ [Fig.~\ref{fig:ds}(b)] and
150$^{\circ}$ [Fig.~\ref{fig:ds}(c)] split the four models into two families 
above  $E_{\gamma}^{lab}~\approx$ 1.5 GeV: in the backward hemisphere,  
both the SL and A models
predict  significantly larger cross sections 
than the two others
(B and C) which embody the off-shell effects.

For the angular distributions [Fig.~\ref{fig:ds}(d), 
\ref{fig:ds}(e), and \ref{fig:ds}(f)], the four models give
similar results at $E_{\gamma}^{lab}$ = 1.0 and 1.45 GeV, while
at the highest energy [$E_{\gamma}^{lab}$ = 2.1 GeV; Fig.~\ref{fig:ds}(f)] the
off-shell treatments produce drastic effects at backward angles.

A striking manifestation of the above behaviors can be seen while
investigating the total cross section (Fig.~\ref{fig:st}). The long
lasting shortcoming of the phenomenological models based on 
effective Lagrangian approaches is significantly cured by the
inclusion of the off-shell effects\footnote{Preliminary data
from ELSA~\cite{BIGSKY} for both differential cross section at about 
2 GeV and the total cross section up to the same energy show trends
similar to those of model B in Figures~\ref{fig:ds}(f) 
and~\ref{fig:st}.}. 
Namely, the total cross-section
does not any more show a diverging behavior above 
$E_{\gamma}^{lab}~\approx$ 1.5 GeV (see also Fig.~5 in Ref.~\cite{SL}).

In the explored phase space region, the excitation functions and angular
distributions for single polarization asymmetries (Fig.~\ref{fig:pol}) show  
significant sensitivity to the off-shell treatments above roughly
1.8 GeV for the $\Lambda$-polarization asymmetry ($P$)
and polarized target asymmetry ($T$). In the case of the 
linearly polarized beam asymmetry ($\Sigma$) the effects are more
drastic. Indeed, above $E_{\gamma}^{lab}~\approx$ 1.6 GeV the off-shell
treatments produce a sign change with sizeable magnitudes around
2 GeV.

The angular distributions for double polarization asymmetries,
at $E_{\gamma}^{lab}$ = 1.45 and 2.1 GeV,
are shown in Fig.~\ref{fig:coxz}. A general trend for these
observables is that significant off-shell effects appear
in the backward hemisphere.  In the case of $C_{x\prime}$
asymmetry, this sensitivity gets attenuated with increasing
photon energy. For the other asymmetry (
$C_{z\prime}$) with circularly
polarized beam, as well as for the two asymmetries 
($O_{x\prime}$ and $O_{z\prime}$) with
linearly polarized beam, the effects are, on the contrary,
enhanced with increasing photon energy. It is worth
noticing that the two models without off-shell treatments
predict almost vanishing values for $O_{x\prime}$ and 
$O_{z\prime}$ asymmetries, while introducing these 
treatments results in a sign change and  sizeable
magnitudes for these asymmetries in the backward hemisphere.

We note that the curves depicted in Figures~\ref{fig:ds}-\ref{fig:coxz} 
split in
two families depending on whether the off-shell effects are included
(models B and C) or not (models SL and A). 

%
%
\subsubsection {Reaction $e~+~p~{\rightarrow}~e'~+~K^+~+~\Lambda$}

The cross section for the electroproduction process is given by
 
\begin{equation}
{d\sigma \over {d\Omega}_K} = {d\sigma}_U + {\epsilon}_L \, 
{d\sigma}_L +
{\epsilon} \, {d\sigma}_P \, {\sin}^2\theta \,{\cos}2\phi +
\sqrt {2{\epsilon}_L(1+{\epsilon})} \, {d\sigma}_I \, 
{\sin}\theta \, {\cos}\phi ,
\label{eq:sigel}
\end{equation}
 
\noindent
with  
 $\theta$ the angle between the outgoing kaon and the virtual 
photon, and $\phi$ the azimuthal angle between the kaon 
production plane and
the electron scattering plane.
Transverse and 
longitudinal polarization parameters 
$\epsilon$ and $\epsilon_L$, respectively, are defined as
 
\begin{equation}
\epsilon = \left[1 -
        2 \frac{|\mbox{\boldmath $p$}_{\gamma}|^2}{p_{\gamma}^2} 
        \tan^2(\frac{\Psi}{2}) \right] \quad , \quad
\epsilon_L = - \frac{p_{\gamma}^2}{p_{\gamma 0}^2}\> \epsilon ,
\end{equation}
 
\noindent
with $\Psi$ the angle between the momenta of the incoming and outgoing
electrons.
 Moreover, ${d\sigma}_U$ is the cross section for an 
unpolarized incident 
photon beam, and the term containing ${d\sigma}_P$ is the asymmetry 
contribution of a transversally polarized beam. The cross section of  
a longitudinally polarized virtual photon is given by ${d\sigma}_L$, 
while ${d\sigma}_I$ contains the interference effects between the 
longitudinal and  transverse components of the beam.

In the figures shown in the remaining of this Section, the electromagnetic 
form factors used are the same as in the SL model (see Subsec. IV.D in
Ref~\cite{SL}).

Figure~\ref{fig:ul} shows the unpolarized component of the 
differential cross section 
$d\sigma_{UL}$  = ${d\sigma}_U + {\epsilon}_L \, {d\sigma}_L$, 
[see Eq.~(\ref{eq:sigel})],
as a function of the momentum transfer. 
All four models reproduce the data equally well.
We note again that models B and C give almost identical results. 
The predictions for different components of the cross section
[Eq.~(\ref{eq:sigel})] are reported in Fig.~\ref{fig:ulip}. The transverse 
component (T) splits also the four curves in the same two families, 
with the off-shell
effects producing significantly smaller values for this observable.
On the contrary, these effects enhance  the longitudinal (L) part.
The transverse-longitudinal (TL) interference term shows similar sensitivities.
Among the observables reported  here, the (L) and (TL) terms are the
only ones to produce the most sizeable differences between the models
SL and A. Finally, the transverse-transverse (TT) interference term shows
rather negligible dependence on the ingredient of the models.

Because of the above predictions on the suppression of the transverse 
component and
the enhancement of the longitudinal one due to the off-shell treatments, 
the ratio $R(t)=d\sigma_L/d\sigma_U$ is an interesting quantity to be 
investigated.
This latter was already found  appealing in the SL approach while examining
the effects of hadrons electromagnetic form factors.
Here, the off-shell treatments have sizeable effect (Fig.~\ref{fig:R}):
the ratio $R(t)$ between $-t=$ 0.5 and 1.0 GeV$^2$ is increased by a factor
of $\approx$ 2 to 4, due to such treatments.
%
%
\subsubsection {Reaction $K^-~+~p~{\rightarrow}~{\gamma}~+~\Lambda$}
The amplitudes of the strangeness photoproduction can be related by crossing 
symmetry~\cite{CROSY} to those of $K^-p$ radiative capture processes
 
\begin{equation}
K^- ~+~ p ~\rightarrow~ \gamma ~+~ \Lambda.
\end{equation}
 
Here, the relevant quantity is the branching ratio defined as
 
\begin{equation}
BR = {{\Gamma(K^-p \rightarrow \gamma \Lambda)} \over
         {\Gamma(K^-p \rightarrow  \mbox{all})}},
\end{equation}
 
\noindent with stopped kaons. 

In Table~\ref{tab:BR}, the results of the four models are compared with the
only available data point. They all agree with the upper bound of
the experimental result. Although in the SL model the presence of the $N7$
resonance was found relevant in reproducing the measured branching
ratio (see Table XI in Ref~\cite{SL}), the off-shell treatments
are not affecting this observable. This may be due to the fact that
here we are dealing only with stopped kaons, and the reported behavior 
might be altered for kaons in flight.

Before ending this section, we wish to make a few comments on some
general features of the findings summarized in Table~\ref{tab:ABC}
and/or depicted in Figures 1 to 7.
%
%

%

\subsection {Comments on free parameters}
The models discussed in this paper embody 12 (model A) to 20 (model C)
free parameters, see Table~\ref{tab:ABC}.
In this subsection, we emphasize that, in spite of rather large
number of free parameters, our approach offers some meaningful
insight into the dynamics of the strangeness electromagnetic
production processes.
%
%
\subsubsection {Coupling constants}
In the fitting procedures, the two main coupling constants 
(Table~\ref{tab:ABC}), $g_{K\Lambda N}$ and $g_{K\Sigma N}$,
have been allowed to vary within their broken $SU(3)$-symmetry limits
~\cite{ade90}.
Given that for the other couplings
we do not dispose of any reliable values or constraints, 
we will discuss their variations, within the corresponding
uncertainties, according to the models ingredients
and/or off-shell treatments. The values referred to concern
models SL, A, B, and C in Table~\ref{tab:ABC}.

\begin{itemize}
\item{\it s-channel: }{no significant variations are observed.}
\item{\it u-channel: }{In going from the SL model to model A,
the couplings of the $L5$ and $S1$ resonances undergo variations of 
factor 2 to 3.
Then the inclusion of off-shell effects (going from model A to B)
brings them back close to their SL model values, stabilizing them for the 
C model. These two consecutive variations might come from the 
observation~\cite{SL}
that in the SL model these two resonances are rather strongly
correlated. This fact, in the absence of any constraint, 
leads to large variations of the $L5$ and $S1$ coupling constants. 
However,
the combined contribution of these resonances to the observables
does not show any drastic variation.
}
\item{\it t-channel: }{Significant variations are noticed comparing
SL model with the other ones. Notice that a spin 5/2 resonance
present in the SL model has been removed in the other models.
The global increase of the {\it t-}channel strengths when discarding
a spin 5/2 resonance is a manifestation of the duality  hypothesis
(the interplay between s- and {\it t-}channel
strengths) in the strangeness sector, as discussed in Ref.~\cite{st96}.
}
\end{itemize}
The above considerations indicate strongly that the underlying
dynamics retained in this work are tightly constrained by the
available data base. Hence, the reported sensitivities to off-shell
treatments are not likely to be altered significantly by the rest of the free 
parameters of the models introduced here.
%
%
%
\subsubsection {Off-shell free parameters}
In obtaining models B and C we have treated the three parameters (X, Y, and Z)
as free ones. As shown in Table~\ref{tab:ABC}, out of six off-shell 
parameters related
to the N7 and L8 resonances, the largest one by far is the Y parameter
for the N7 resonance. The Z parameter related to this latter resonance
comes out to be compatible with zero. Moreover, all three parameters of 
the N7 resonance are stable upon comparing B and C models.

Notice that one of the main motivations in introducing the off-shell
effects is to cure an undesirable increase in the predicted photoproduction
total cross-section above roughly 1.5 GeV. By examining the non-pole part 
$P^{NP}_{ij}$ of the amplitudes  in Appendix~\ref{app:invamp}, one 
finds~\footnote{
From Eqs.~\ref{eq:p1jnp} and~\ref{eq:p2jnp} in 
Appendix~\ref{app:invamp}, we see that only three of the non-pole 
coefficients ($P_{11}^{NP}$, $P_{21}^{NP}$, and $P_{23}^{NP}$) depend
on the $s$ variable and that this dependence is linear. We write hence
$P_{ij}^{NP}$~=~$a_{ij}s$~+~$b_{ij}$ , where the coefficients $a_{ij}$ and
$b_{ij}$ are functions of only off-shell parameters and baryons masses. 
Then one can readily derive the following expressions:

$a_{11} \propto (\widetilde Z-1/2)\> (2\widetilde Y-1)$; 
$a_{21} \propto (\widetilde Z-1/2) + \widetilde X\>
[\>1-2 \widetilde Z\> (M_Y/M_R +2) ]$; 
$a_{23} \propto \widetilde X\> (\widetilde Z-1/2)$.

\noindent
All these coefficients vanish at $\widetilde X=0$ and 
$\widetilde Z=1/2$ (i.e., $X=-1/2$ and $Z=0$).
}
that for the off-shell parameters $X \neq -1/2$ and $Z \neq 0$, 
there are contributions to the invariant amplitudes which rise linearly 
as a function of the
{\it s-}variable (observe that $Y$ does not participate in this
matter). Hence, the cross section increase stated above might be 
due to the $X \neq -1/2$ and $Z \neq 0$ values, 
as obtained from the present minimizations (Table~\ref{tab:ABC})
exploiting the available data.

The authors of Ref.~\cite{Ben89} have discussed extensively different
"choices" of these free parameters, and especially fixing two of them,
as reported in the literature~\cite{nat80}. 
They conclude that there is no physical
basis to attribute fixed values to any of these off-shell parameters.

However, to numerically estimate the consequences of eliminating the
undesirable {\it s}-dependent terms by imposing $X = -1/2$ and $Z = 0$,
we have performed a minimization within the context of 
model B.
The results for the coupling constants and the only adjusted off-shell 
parameter ($Y$)
are given in Table~\ref{tab:ABC} as
model D. We see that the only significant variation compared to
model B concerns the $Y$-parameter.
Notice that for model B we had already $Z \approx 0$. Hence, decreasing
the magnitude of the $X$ parameter by roughly a factor of 2 (between models
B and D) leads to an increase of about 20\% of the magnitude of
the $Y$-parameter.
In Figure~\ref{fig:XZ} the photoproduction total cross section and 
the electroproduction ratio $R(t)=d\sigma_L/d\sigma_U$ 
are depicted for both B and D models.
In both cases the results from the two models are quite close and
the photoproduction total cross section comes out to give slightly
higher values using the {\it ad hoc} fixed values for $X$ and $Z$ 
[Fig.~\ref{fig:XZ}(a)]. 
Other observables discussed in this paper show similar behaviors
while comparing models B and D.  
The closeness of the predictions for the observables can be 
understood by noticing that the $Z$ parameter in
model B has a value compatible with zero, and the contributions
due to the $Y$-dependent terms dominate numerically over those
coming from the $Z$-dependent ones.

In the case of pion photoproduction, the RPI-group~\cite{dav91}
found that imposing $X = -1/2$ and $Z = 0$ leads to a significant
increase of the $\chi^2$. This is not the case with the present
investigation (Table~\ref{tab:ABC}). The reason is that
the pion photoproduction was studied in the $\Delta_{33}$ resonance
region, where the reaction mechanism is dominated by this spin-3/2
resonance, while in the case of strangeness production none of the
resonances has a paramount role. Moreover, we recall that
the model B (C) studied
here contains one (two) spin-3/2 resonance and five spin-1/2 resonances.

To our knowledge, there are {\it a priori} no bounds on the values
of the off-shell parameters. 
However, remembering that the off-shell freedom comes in
only from the non-pole terms, and that the principal contribution 
from a given resonance must correspond to the propagation of its
proper spin, we expect that the corresponding non-pole parts 
might not dominate the pole part. This would
give reasonable upper-bound to which values $X$, $Y$, and $Z$ may take.
This expectation was verified in the case of the models reported here.

Finally, in the case of $L8$ hyperonic resonance (model C), 
very small values of the off-shell parameters, as well as those of
coupling constants resulting from the minimization
endorse our previous affirmations: contributions from this resonance 
are not required by the existing data base, and the smallness of the 
relevant free parameters explains why its inclusion in the underlying
dynamics does not have significant consequences, neither on the 
$\chi^2$ nor on the predicted observables.

%
  

\section{SUMMARY AND CONCLUSIONS}

In the present article, focused on the electromagnetic production of
strangeness, we have been concerned with the improvement on the
effective hadronic Lagrangian approaches by incorporating the correct 
spin-3/2 resonances propagators and what is called off-shell effects 
entering the vertices connected to these resonances.

The work presented here allows us to preserve the gauge invariance of 
the formalism,
to ensure that each propagator associated with a spin-3/2 exchanged 
baryon has an inverse, and 
to include simultaneously both $N^*$ and
$Y^*$ spin-3/2 resonances.

Applying our approach to the $K\Lambda$ channels observables
investigated in Ref~\cite{SL}, we have emphasized that 
the photo- and electro-production of 
$K^+ \Lambda$ observables show significant sensitivity to 
the off-shell effects, while these effects do not lead to measurable
manifestations in the 
$K^-p$ radiative capture branching ratio with stopped kaons.

The numerical results reported here are of course heavily based
on the existing data. Given the inconsistencies~\cite{ade90} within the
present data, the dynamical content of the models reported here 
will very likely evolve with the forthcoming high quality data from
several  experiments, both ongoing and planned. Hence the presented
numerical results should be considered as guidelines for relative 
effects. The efforts in refining the phenomenological approaches
are then meant to provide us with appropriate tools to interpret 
the upcoming data. 

Applying the formalism derived in this paper to the available data-base,
our results show that the photoproduction data, especially polarization
asymmetries, are crucial in pinning down the role of off-shell effects. 
Once these effects are under control, the electroproduction channel can be
investigated in studying the electromagnetic form factors of the baryons, 
kaon and their resonances. These conclusions were reached for the 
$K \Lambda$ channels and we are currently investigating the complementary
$K \Sigma$ processes.

There are a few items not discussed in our
investigation here for which  some comments may be due.
 
In none of the reported approaches (including this work) the constructed
amplitudes embody unitarity.
Recently, there has been some attempts to 
unitarize the amplitude in this process. Lu {\it et al.}~\cite{lu95} 
have performed a {\it ``feasibility study''} within a chiral color
dielectric model adopting a simple two-channel case and using
the $K^+ N$ phase shifts to approximate the $K^+ \Lambda$ elastic
scattering in the final state. 
Kaiser {\it et al.}~\cite{kai97} have developed an SU(3) chiral dynamics with
an effective coupled-channel potential. This {\it s}-wave approximation 
approach is  limited to the near threshold region. These works put
forward some indications on the effects from the unitarization, but they  
do not offer a definitive conclusion about the importance of the final
state interactions.  

When several final channels are
taken into account, to be realistic, a complete unitarization procedure becomes
beyond our current capacity. Note also that there has not been any unique
way the unitarization should be carried on. We thus believe and hope that, 
since the finite widths of the resonances are included, some parts 
contributing towards unitarization have been effectively included in the models
discussed in this paper. 

Before ending this Section, we wish to discuss briefly two recent and more
fundamental approaches applied to some of the processes investigated 
in this paper.

(i) {\it Chiral Perturbation Theory (CHPT)}: 
This approach {\it limited to the threshold region},  
incorporates the
coupling to baryon multiplets, as has recently been done in Ref.~\cite{CHPT} 
putting
more emphasis on the consequence from the strict chiral symmetry in the
construction of Lagrangian. Here we simply note that the predictions of
this model reproduce reasonably the low energy total photoproduction 
cross sections and the recoil polarization $P$, and are qualitatively 
consistent with our results. To establish the quality of the {\it CHPT} 
predictions 
extended to three flavors and including baryons, more data near threshold 
are needed.   

(ii) {\it Quark Models}:
These models (upon adopting some type of
chiral quark model, for example Ref.~\cite{ZPL}) can predict 
certain observables 
with less number of free parameters than the Effective Hadronic Lagrangian
(EHL) approaches, given the consequence of the 
difference in the
underlying quark models is well within some controllable limit. Right now
{\it the electroproduction process is rather hard to deal with} by the existing
quark model approaches to the strangeness production.

We observe that in the current stage of development, the Quarks models
and the EHL approaches are somewhat 
complementary. With available data and upon minimization, EHL can supply the
values of combined coupling constants like $G_1$ and $G_2$ 
in our present approach.
Then, the thus obtained amplitudes are able to predict yet unmeasured
observables. Once those observables are measured, they will serve in
constraining the hadronic coupling values. They are then decomposed into pure 
hadronic and strong-electromagnetic parts, to constrain the underlying 
sub-hadronic dynamics (selecting certain quark models, for example). 

Some quantities are present in one approach which are absent in the other 
(like the off-shell effects). We still lack a 
microscopic covariant approach within quark models which could, in principle,
answer questions related to these aspects.

Concluding, the complementarity between the Effective Lagrangian approach and 
other promising investigations~\cite{lu95,kai97,CHPT,ZPL}, provide us with
powerful theoretical means to interpret the copious and high quality data
to come.


\section*{Acknowledgments}
We would like to thank Jean-Christophe David, Zhenping Li 
and Nimai Mukhopadhyay for fruitful
discussions, and Dietmar Menze and Reinhard Schumacher for helpful
exchanges on the experimental results and projects.
\newpage

\appendix

\section{}
\protect\label{app:cgln}

\centerline{{\bf CGLN amplitudes}}

The well-known Chew, Goldberger, Low and Nambu ($CGLN$) amplitudes
entering into the expressions of the photo- and electro-production observables
(see for example Ref.~\cite{SL}) are related to the ${\cal{A}}_j$ 
invariant functions as follows:

\begin{eqnarray}
{\cal{F}}_1 & = & ({\sqrt s} - M_p) {\cal{A}}_1
- p_{\gamma} {\cdot}p_p {\cal{A}}_3 - 
p_{\gamma} {\cdot}p_Y {\cal{A}}_4
- p^2_{\gamma}\;{\cal{A}}_5 ,  \\ [10pt]
{\cal{F}}_2 & = & \frac{\vert {\bbox{p}}_{\gamma}\vert \vert
 {\bbox{p}}_{K}\vert}
{(E_p + M_p)(E_Y + M_Y)}
\bigg[({\sqrt s} + M_p) {\cal{A}}_1
+ p_{\gamma} {\cdot}p_p {\cal{A}}_3 + p_{\gamma} {\cdot}p_Y 
{\cal{A}}_4 + p^2_{\gamma}\;{\cal{A}}_5\bigg] , \\ [10pt]
{\cal{F}}_3 & = & \frac{\vert {\bbox{p}}_{\gamma}\vert \vert
 {\bbox{p}}_{K}\vert}
{(E_p + M_p)}
\bigg[- 2 p_{\gamma} {\cdot}p_p {\cal{A}}_2
+ ({\sqrt s} + M_p) {\cal{A}}_4 +
p^2_{\gamma}\;{\cal{A}}_6 \bigg] ,\\ [10pt]
{\cal{F}}_4 & = & \frac{{\vert {\bbox{p}}_{K}\vert}^2}
{(E_Y + M_Y)}
\bigg[2 p_{\gamma} {\cdot}p_p {\cal{A}}_2
+ ({\sqrt s} - M_p) {\cal{A}}_4
-  p^2_{\gamma}\;{\cal{A}}_6 \bigg] , \\ [10pt]
{\cal{F}}_5 & = & \frac{{\vert {\bbox{p}}_{\gamma}\vert}^2}
{(E_p + M_p)}
\bigg[- {\cal{A}}_1 + 2 p_{\gamma} {\cdot}p_Y {\cal{A}}_2 +
({\sqrt s} + M_p)({\cal{A}}_3 - {\cal{A}}_5) +
(p_{\gamma} {\cdot}p_Y 
- p_{\gamma} {\cdot}p_p - p^2_{\gamma}) {\cal{A}}_6\bigg] , \\ [10pt]
{\cal{F}}_6 & = & \frac{\vert {\bbox{p}}_{\gamma}\vert \vert
 {\bbox{p}}_{K}\vert}
{(E_Y + M_Y)}
\bigg[- 2 p_{\gamma} {\cdot}p_Y {\cal{A}}_2 + ({\sqrt s} - M_p)
{\cal{A}}_3 - 
(p_{\gamma} {\cdot}p_Y 
- p_{\gamma} {\cdot}p_p - p^2_{\gamma}) {\cal{A}}_6 - \\
 &  & \hskip 2.3 truecm {1\over E_p + M_p} \bigg\{
 p_{\gamma 0} {\cal{A}}_1 + p_{\gamma} {\cdot}p_p {\cal{A}}_3  +
p_{\gamma} {\cdot}p_Y {\cal{A}}_4 + p_{\gamma 0} ({\sqrt s} + M_p) 
{\cal{A}}_5 \bigg\}\bigg] .
\end{eqnarray}

The only differences with the relations given in Appendix D of Ref~\cite{SL}
appear in the amplitudes ${\cal{F}}_5$ and ${\cal{F}}_6$ (contributing only in 
the electroproduction observables) where we have the following extra term
inside the braces: $- (p_{\gamma} {\cdot}p_p + p^2_{\gamma}) {\cal{A}}_6$.

\newpage

\section{}
\protect\label{app:invamp}

\centerline{{\bf Invariant amplitudes from an {\it s-}channel spin 3/2 
resonance}}

Here we present the concrete form for the invariant amplitudes 
decomposed into the pole (P) and non-pole (NP) parts as discussed in 
Eqs.~(\ref{eq:ajphoto}) to (\ref{eq:decomp}).  The calculation has been done 
both manually, and by using MAPLE to confirm the validity of the 
former.

To begin, we first introduce several coefficients as well as a few 
Lorentz scalar products which enter the expressions for the amplitudes.

\begin{eqnarray}
{A} &=&  - \,{\displaystyle \frac {1}{6\,M_R^ 2}}\,
          (\, M_Y^2 + M_R^2 - M_K^2 - M_R M_Y \,),\\
{B}(\,{\widetilde Z}\,) &=&    \,{\displaystyle 
                               \frac{1 - 2 \widetilde Z}{6\, M_R^2}}, \\
{C} &=& {\displaystyle \frac {1}{12 \,M_R^{2}\,
           (\,p_{\gamma}{\cdot}p_Y - p_{\gamma}{\cdot}p_p\,)}}
\left[ \,
2\,M_R\,M_p\,M_Y - (\,M_Y^{2}
 + M_R^{2} - M_K^{2}\,)\,(\,2\,M_p - 3\,M_R\,)\, \right], \\
{D}(\,{\widetilde X}, {\widetilde Z}\,) &=& 
{\displaystyle \frac {1}{12 \,M_R^{2}\,
(\,p_{\gamma}{\cdot}p_Y - p_{\gamma}{\cdot}p_p\,)}}
\left[ \,(\,2\,M_p - M_R\,) 
    - 2\,(\,M_R + 2\,M_Y + 2\,M_p\,)\,{\widetilde Z}
    \right. \cr 
    &&\mbox{\hspace{185pt}} -\left. 2\,M_R\,{\widetilde X}
     +4\,(\,M_Y + 2\,M_R\,)\,{\widetilde X}\,{\widetilde Z}
    \, \right], \\
{E} &=&   \,{\displaystyle \frac {1}{12\,M_R}}\,
          \left[ \, M_K^2 - \,(\, M_Y + M_R \,)^2 \,\right], \\
{F}(\,{\widetilde X}, {\widetilde Z}\,) &=& 
    {\displaystyle \frac {1}{12\,M_R^2}}\, 
        [\,M_R - 2\, M_R\, {\widetilde Z} - 2\, M_R\, {\widetilde X}
         + 4\,(\,M_Y+ 2\,M_R\,)\,{\widetilde X}\,{\widetilde Z} \,] .
\end{eqnarray}

The dot products are given by 

\begin{equation}
p_{\gamma}{\cdot}p_p = \frac{1}{2}(s-M_p^2-p_{\gamma}^2) \quad , \quad
p_{\gamma}{\cdot}p_Y = \frac{1}{2}(M_Y^2+p_{\gamma}^2-u),
\end{equation}

\noindent
Using the relation 
$s + t + u = M_p^2 + M_Y^2 + M_K^2 + p_{\gamma}^2$, 
we obtain 
\begin{equation}
p_{\gamma}{\cdot}p_Y - p_{\gamma}{\cdot}p_p = 
                           \frac{1}{2}(t-M_K^2-p_{\gamma}^2), 
\end{equation} 

\noindent With this preparation above we first present the quantities 
$P_{1j}^{P, NP}$ ($j=1,...4$) for the photoproduction coming from the 
$G_1$ coupling

\begin{eqnarray}
\label{eq:p1j}
P^P_{11} &=&   \left( \! \,{\displaystyle \frac {1}{
6}}\,{\displaystyle \frac {M_p^{2}}{M_R^{2}}} - 
{\displaystyle \frac {1}{3}}\,{\displaystyle \frac {M_p}{
M_R}} - {\displaystyle \frac {1}{2}}\, \!  \right) \,M_Y^{2}
 +  \left( \! \, - \,{\displaystyle \frac {1}{6}}\,
{\displaystyle \frac {M_p^{2}}{M_R}} - {\displaystyle 
\frac {2}{3}}\,M_p - {\displaystyle \frac {1}{2}}\,M_R
\, \!  \right) \,M_Y \cr
 & & \mbox{} +  \left( \! \, - \,{\displaystyle \frac {1}{3}} - 
{\
\displaystyle \frac {1}{6}}\,{\displaystyle \frac {M_K^{2}
}{M_R^{2}}}\, \!  \right) \,M_p^{2} +  \left( \! \, - 
\,{\displaystyle \frac {1}{3}}\,M_R + {\displaystyle \frac {
1}{3}}\,{\displaystyle \frac {M_K^{2}}{M_R}}\, \! 
 \right) \,M_p + {\displaystyle \frac {1}{2}}\,{t},
\mbox{\hspace{51pt}} \nonumber \\
P^P_{12} &=& 1, \nonumber \\
P^P_{13} &=&  {\displaystyle \frac {1}{3}}\,{\displaystyle 
\frac {M_Y^{2}\,M_p}{M_R^{2}}} +  \left( \! \, - 
\,{\displaystyle \frac {1}{3}}\,{\displaystyle \frac {M_p}{
M_R}} - 1\, \!  \right) \,M_Y +  \left( \! \,
{\displaystyle \frac {1}{3}} - {\displaystyle \frac {1}{3}}\,
{\displaystyle \frac {M_K^{2}}{M_R^{2}}}\, \!  \right) 
\,M_p, \nonumber \\
P^P_{14} &=&   - (M_p + M_R), \nonumber \\
%
\end{eqnarray}
\begin{eqnarray}
\label{eq:p1jnp}
P^{NP}_{11} &=&  {\displaystyle \frac {2}{3}}\,
{\displaystyle \frac {(\,{s} + M_Y\,M_p + 2\,M_R\,
M_Y + 2\,M_R\,M_p\,)\,{\widetilde Y}\,{\widetilde Z}}{M_R
^{2}}} \cr
 & & \mbox{} - {\displaystyle \frac {1}{3}}\,{\displaystyle 
\frac {(\,{s} - M_Y^{2} + M_R\,M_Y + M_R\,
M_p + M_K^{2}\,)\,{\widetilde Y}}{M_R^{2}}} \cr
 & & \mbox{} + {\displaystyle \frac {1}{3}}\,{\displaystyle 
\frac {(\, - {s} - 2\,M_R\,M_Y + M_p^{2} - 2\,
M_R\,M_p\,)\,{\widetilde Z}}{M_R^{2}}} \cr
 & & \mbox{} - {\displaystyle \frac {1}{6}}\,{\displaystyle 
\frac { - {s} + M_Y^{2} - M_R\,M_Y + M_p^{2}
 - 2\,M_R\,M_p - M_K^{2}}{M_R^{2}}},
\mbox{\hspace{26pt}} \nonumber \\
P^{NP}_{12} &=&  0, \nonumber \\
P^{NP}_{13} &=&  {\displaystyle \frac {4}{3}}\,
{\displaystyle \frac {(\,M_Y + 2\,M_R\,)\,{\widetilde Y}\,
{\widetilde Z}}{M_R^{2}}} - {\displaystyle \frac {2}{3}}\,
{\displaystyle \frac {{\widetilde Y}}{M_R}} + {\displaystyle 
\frac {2}{3}}\,{\displaystyle \frac {(\, - M_Y + M_p - 
2\,M_R\,)\,{\widetilde Z}}{M_R^{2}}} \cr
 & & \mbox{} - {\displaystyle \frac {1}{3}}\,{\displaystyle 
\frac {M_p - M_R}{M_R^{2}}},\mbox{\hspace{236pt}} \nonumber \\
P^{NP}_{14} &=&  0. \nonumber \\
\end{eqnarray}
Those coming from the $G_2$ coupling, viz.  $P_{2j}^{P, NP}$ 
(j=1,...4) are 

\begin{eqnarray}
\label{eq:p2j}
P^P_{21} &=& - E \, (\,M_R^2 - M_p^2\,), \nonumber \\
P^P_{22} &=&  {\displaystyle \frac {1}{2}}\, (\,M_p - M_R\,), \nonumber \\
P^P_{23} &=&   - \,{\displaystyle \frac {1}{6}}\,
{\displaystyle \frac {(\,M_Y + 2\,M_R\,)\,(\,M_p
 - M_R\,)\,M_Y}{M_R}} + {\displaystyle \frac {1}{2
}}\,M_p^{2} 
 + {\displaystyle \frac {1}{6}}\,{\displaystyle 
\frac {(\,M_K^2 - M_R^2\,)\,(\,M_p + 2\,M_R\,)}{M_R}} -
{\displaystyle \frac {1}{2}}\,{t}\mbox{\hspace{1pt}}, \nonumber \\
P^P_{24} &=&   - \,{\displaystyle \frac {1}{2}}\,
                     (\,M_p^{2} - M_R^{2}\,), \nonumber \\
\end{eqnarray}
\begin{eqnarray}
\label{eq:p2jnp}
P^{NP}_{21} &=& - E  - {F}(\,{\widetilde X}, {\widetilde Z}\,)
\,(\,{s} - M_p^{2}\,), \nonumber \\
P^{NP}_{22} &=&  0, \nonumber \\
P^{NP}_{23} &=&  {\displaystyle \frac {2}{3}}\,
{\displaystyle \frac {(\, - {s} + M_Y\,M_p - 2\,M_R
\,M_Y + 2\,M_R\,M_p\,)\,{\widetilde X}
\,{\widetilde Z}}{M_R^{2}}} \cr
 & & \mbox{} - {\displaystyle \frac {1}{3}}\,{\displaystyle 
\frac {(\, - {s} + M_Y^{2} - M_R\,M_Y + M_R\,
M_p - M_K^{2}\,)\,{\widetilde X}}{M_R^{2}}} - 
{\displaystyle \frac {1}{3}}\,{\displaystyle \frac {(\, - M_Y 
+ M_p\,)\,{\widetilde Z}}{M_R}} \cr
 & & \mbox{} + {\displaystyle \frac {1}{6}}\,{\displaystyle 
\frac {M_p - 2\,M_R}{M_R}}, \nonumber \\
P^{NP}_{24} &=&  {\displaystyle \frac {1}{2}}. \nonumber \\
\end{eqnarray}
The parts for $j= 1,...,4$ contributing to the electroproduction  
[see Eq.~(\ref{eq:decomp})] are as follows:

 - those related to coupling $G_1$
 
\begin{eqnarray}
\label{eq:r1j}
R^P_{11} &=&  {A} \quad , \quad
R^P_{12}  = {\displaystyle 
 \frac {2\,A - 1}{2\,(\,p_{\gamma} {\cdot} p_Y - 
                        p_{\gamma} {\cdot} p_p \,)}}, \nonumber \\
R^P_{13} &=&  R^P_{14} = 0,  \nonumber \\
      \\
R^{NP}_{11} &=&  {B}(\,{\widetilde Z}\,) \quad , \quad
R^{NP}_{12}  = {\displaystyle \frac {{B}(\,{\widetilde Z}\,)}
   {p_{\gamma} {\cdot} p_Y - p_{\gamma} {\cdot} p_p}}, \nonumber \\
R^{NP}_{13} &=&  R^{NP}_{14}  =  0, \nonumber \\
\nonumber
\end{eqnarray}

- those related to coupling $G_2$

\begin{eqnarray}
\label{eq:r2j}
R^P_{21} &=&  {E} \quad , \quad
R^P_{22}  =  {C}, \nonumber \\
R^P_{23} &=&  - 2\,{A} \quad , \quad
R^P_{24} =  {\displaystyle - \frac {1}{2}}, \nonumber \\
      \\
R^{NP}_{21} &=&  {F}(\,{\widetilde X}, {\widetilde Z}\,) \quad , \quad
R^{NP}_{22} = {D}(\,{\widetilde X}, {\widetilde Z}\,), \nonumber \\ [2pt]
R^{NP}_{23} &=&  - 2\,{B}(\,{\widetilde Z}\,) \quad , \quad
R^{NP}_{24} =  0. \nonumber \\
\nonumber
\end{eqnarray}

For $j=5,6$ (contributing solely to the electroproduction) the $E_{ij}$ 
coefficients coming from $G_1$ are 

\begin{eqnarray}
\label{eq:e1-56}
E^P_{15} &=& - 2 \, {A}\,(\, M_R + M_p \,) \quad , \quad
E^P_{16}  =  {\displaystyle 
\frac {2 \, A\, p_{\gamma} {\cdot} p_p - p_{\gamma} {\cdot} p_Y }
   {{p_{\gamma} {\cdot} p_Y - p_{\gamma} {\cdot} p_p}}}, \nonumber \\
   \\
E^{NP}_{15} &=& - \,{\displaystyle \frac {1}{3}}\,
{\displaystyle \frac {M_p}{M_R^{2}}} + {\displaystyle \frac {2}{3}}\,
{\displaystyle \frac {(\, - M_Y - M_R + M_p\,)\,
{\widetilde Z}}{M_R^{2}}} - {\displaystyle \frac {1}{3}}\,
{\displaystyle \frac {{\widetilde Y}}{M_R}} + {\displaystyle 
\frac {2}{3}}\,{\displaystyle \frac {(\,M_Y + 2\,M_R\,)
\,{\widetilde Y}\,{\widetilde Z}}{M_R^{2}}}, \nonumber \\ 
E^{NP}_{16} &=&  {\displaystyle 
\frac {2\,B(\,{\widetilde Z}\,)\, p_{\gamma} {\cdot} p_p }
{{p_{\gamma} {\cdot} p_Y - p_{\gamma} {\cdot} p_p}}}. \nonumber \\
\nonumber
\end{eqnarray}

\noindent
while those coming from $G_2$ are

\begin{eqnarray}
\label{eq:e2-56}
E^P_{25} &=& 2\,{A}\,p_{\gamma} {\cdot} p_p \quad , \quad 
E^P_{26}  =  2\,{C}\,p_{\gamma} {\cdot} p_p, \nonumber \\ 
    \\
E^{NP}_{25} &=& 2\,B(\,{\widetilde Z}\,)\,p_{\gamma} {\cdot} p_p
                                          \quad , \quad
E^{NP}_{26}  =  2\,D(\,{\widetilde X},{\widetilde Z}\,)
                  \,p_{\gamma} {\cdot} p_p.  \nonumber \\
\nonumber
\end{eqnarray}


\newpage

\section{}

\protect\label{app:nooff}

\centerline{{\bf Vertices adopted by Adelseck {\it et al.}}}

\bigskip

Here we show how the vertices Eqs.~(\ref{eq:vertk}) and 
(\ref{eq:vert32}) may be reduced, by some assumption and 
approximation, to the ones used by Adelseck {\it et al.} \cite{Adel1}.

As discussed in \cite{Ben89}, the propagator adopted by Adelseck {\it 
et al.} Eq.~(\ref{eq:prop32Adel}) may be rewritten ({\it in the limit 
of zero width}) as
\begin{equation}
P_{\mu \nu}^A(q) =\frac{{q}\hskip -0.19 cm / + \sqrt{s}}{s -M^2_R}
{\cal {P}}_{\mu \nu}^{3/2}(q),
\end{equation}

\noindent where ${\cal {P}}_{\mu \nu}^{3/2}(q)$ is the projection 
operator for spin $3/2$ states.  Thus this choice of the propagator 
cuts out the propagation of spin 1/2 states.  With this the scattering 
amplitude Eq.~(\ref{eq:mfi}) reads

\begin{equation}
M_{fi}^{(s)} =
\overline U_Y({\hbox{\boldmath{$p$}\unboldmath}}_{Y})
 {\cal{V}}^{\mu}(K Y R)\ \frac{{q}\hskip -0.19 cm / +\sqrt{s}}{s - M^2_R+ 
i\Gamma_RM_R} 
{\cal {P}}_{\mu \nu}^{3/2}(q)
 {\cal{V}}^{\nu}(R p \gamma)\ U_p({\hbox{\boldmath{$p$}\unboldmath}}_{p}).
\label{eq:mfiAdel}
\end{equation}

For an on-mass-shell positive energy resonance the spin 3/2 projection 
operator may be written as

\begin{equation}
{\cal {P}}_{\mu \nu}^{3/2}(q) = 
\sum U_{\mu}({\hbox{\boldmath{$q$}\unboldmath}})
\overline U_{\nu}({\hbox{\boldmath{$q$}\unboldmath}}), 
\label{eq:projection}
\end{equation}

\noindent where the summation is implied over the spin eigenstates.

By assuming that the propagating spin 3/2 resonance is approximately
on-mass-shell, and in a positive energy state, we find

\begin{equation}
M_{fi}^{(s)} \approx
\sum \overline U_Y({\hbox{\boldmath{$p$}\unboldmath}}_{Y})
 {\cal{V}}^{\mu}(K Y R)U_{\mu}({\hbox{\boldmath{$q$}\unboldmath}})\ 
 \frac{\sqrt{s} + M_R}{s-M^2_R +
i\Gamma_RM_R}\overline U_{\nu}({\hbox{\boldmath{$q$}\unboldmath}})
 {\cal{V}}^{\nu}(R p \gamma)\ U_p({\hbox{\boldmath{$p$}\unboldmath}}_{p}).
\label{eq:mfiAdel2}
\end{equation}

\noindent In the above expression  the equation

\begin{equation}
({q}\hskip -0.19 cm / -M_R)U_{\mu}({\hbox{\boldmath{$q$}\unboldmath}}) = 0,
\end{equation}

\noindent has been used.  So by this assumption (or approximation), we 
have only to find out the structure of (by suppressing the index for 
spin eigenstates) the following matrix elements

\begin{equation}
\overline U_Y({\hbox{\boldmath{$p$}\unboldmath}}_{Y})
 {\cal{V}}^{\mu}(K Y R)U_{\mu}({\hbox{\boldmath{$q$}\unboldmath}}),
\label{eq:vertkAdel}
\end{equation}

\noindent and

\begin{equation}
\overline U_{\nu}({\hbox{\boldmath{$q$}\unboldmath}})
                                                 {\cal{V}}^{\nu}(R p \gamma)
        \ U_p({\hbox{\boldmath{$p$}\unboldmath}}_{p}).
\label{eq:vert32Adel}
\end{equation}

First, by disregarding the off-shell freedom, the $K Y R$ vertex 
in Eq.~(\ref{eq:vertk}), upon sandwiched between two spinors 
Eq.~(\ref{eq:vertkAdel}), becomes

\begin{equation}
\overline U_Y({\hbox{\boldmath{$p$}\unboldmath}}_{Y})
{\cal {V}}^{\mu}(K Y R)U_{\mu}({\hbox{\boldmath{$q$}\unboldmath}}) = 
\frac{g_{K Y R}}{M_K}p^{\mu}_Y\overline U_Y
({\hbox{\boldmath{$p$}\unboldmath}}_{Y})
U_{\mu}({\hbox{\boldmath{$q$}\unboldmath}}),
\end{equation}

\noindent which results from

\begin{equation}
q^{\mu}U_{\mu}({\hbox{\boldmath{$q$}\unboldmath}}) = 
      (p_K + p_Y)^{\mu}U_{\mu}({\hbox{\boldmath{$q$}\unboldmath}}) = 0.
\end{equation}

\noindent This is a consequence from one of the constraints for spin 
3/2 field $R_{\mu}$, recall Section II.A:
$\partial^{\mu}R_{\mu}= 0$.

\noindent Thus by introducing $\tilde g_{K Y R}$ through

\begin{equation}
\frac{\tilde g_{K Y R}}{M_R} \equiv \frac{g_{K Y R}}{M_K},
\end{equation}

\noindent we may identify the $K Y R$ vertex of Adelseck {\it et 
al.} as

\begin{equation}
{\cal {V}}^{\mu}(K Y R) \approx \frac{\tilde g_{K Y R}}{M_R}
p^{\mu}_Y.
\end{equation}

We now look at the $Rp\gamma$ vertex whose matrix element is defined 
in Eq.~(\ref{eq:vert32Adel}).  With no off-shell freedom implemented, 
the vertex Eq.~(\ref{eq:vert32}) reads
\begin{equation}
{\cal {V}}^{\nu}(Rp\gamma) = \left[ \frac{eg_1}{2M_p}(\epsilon^{\nu} 
{p} \hskip -0.16 cm / _{\gamma}
-p^{\nu}_{\gamma}{\epsilon} \hskip -0.16 cm / ) + \frac{eg_2}{4M^2_p}(\epsilon 
\cdot p_pp_{\gamma}^{\nu}
 - p_{\gamma} \cdot p_p \epsilon^{\nu})\right]i\gamma^5.
\label{eq:vert32noff}
\end{equation}

\noindent The second term in the large bracket can be handled quite 
easily: even without taking its matrix element, we can simply define 
the coupling constant $g_b$ through

\begin{equation}
\frac{g_b}{(M_R + M_p)^2} \equiv \frac{eg_2}{4M^2_K}. 
\end{equation}

 \noindent Next, to find $g_a$ we take the matrix element of the first 
 term in the large bracket of Eq.~(\ref{eq:vert32noff}), that is 
 proportional to $g_1$. This reads

\begin{equation}
 i\frac{eg_1}{2M_p}\overline U_{\nu}({\hbox{\boldmath{$q$}\unboldmath}})
      (\epsilon^{\nu}{p} \hskip -0.16 cm / _{\gamma} -
p_{\gamma}^{\nu}{\epsilon} \hskip -0.16 cm / )\gamma^5 
       U({\hbox{\boldmath{$p$}\unboldmath}}_{p}).
\label{eq:g1part}
\end{equation}

Then we exploit the following relations

\begin{eqnarray}
p_{\gamma} = q- p_p,\\
{p} \hskip -0.16 cm / _{\gamma} \gamma^5 
   U({\hbox{\boldmath{$p$}\unboldmath}}_{p}) = -M_p\gamma^5
   U({\hbox{\boldmath{$p$}\unboldmath}}_{p}),\\
\overline U_{\nu}({\hbox{\boldmath{$q$}\unboldmath}}){q}\hskip -0.19 cm / = 
                  \overline U_{\nu}({\hbox{\boldmath{$q$}\unboldmath}})M_R.
\end{eqnarray}   

\noindent
Then Eq.~(\ref{eq:g1part}) may be rewritten as

\begin{equation}
eg_1\frac{(M_R +M_p)}{2M_p}
\overline U_{\nu}({\hbox{\boldmath{$q$}\unboldmath}})\left[\epsilon^{\nu} -
\frac{p_{\gamma}^{\nu}}{M_R+M_p}{\epsilon} \hskip -0.16 cm /  \right]
i\gamma^5 U({\hbox{\boldmath{$p$}\unboldmath}}_{p}).
\label{eq:gacoup}
\end{equation}

Thus by defining $g_a$ through 

\begin{equation}
\frac{g_a}{M_R+M_p} \equiv \frac{eg_1}{2M_p},
\end{equation}

\noindent the Eq.~(\ref{eq:gacoup}) reads

\begin{equation}
\overline U_{\nu}({\hbox{\boldmath{$q$}\unboldmath}})g_a\left[\epsilon^{\nu}
-\frac{p_{\gamma}^{\nu}}{M_R+M_p}{\epsilon} \hskip -0.16 cm / \right]
          U({\hbox{\boldmath{$p$}\unboldmath}}_{p}).
\end{equation}

Then everything put together, the $Rp\gamma$ vertex becomes

\begin{equation}
{\cal V}^{\nu}(R p \gamma) \approx i\ \left[
g_a \left(\epsilon^{\nu}
- \frac{p_{\gamma}^{\nu}{\epsilon \hskip -0.16 cm / }}
{M_R +M_p}\right)
+g_b \frac{1}{(M_R +M_p)^2}(\epsilon 
{\cdot}p_p p_{\gamma}^{\nu} -p_\gamma{\cdot}p_p \epsilon^{\nu})
\right] \gamma^5.
\label{eq:vn32Adel1}
\end{equation} 

A trouble with this form is that it does not respect gauge invariance.  
Thus in \cite{Adel1} the replacement $M_R \to \sqrt{s}$ was made, 
which eventually leads to Eq.~(\ref{eq:vn32Adel}).
%
%

%
%
\squeezetable
\begin{table}
\caption{Exchanged particles, coupling constants, and off-shell
parameters ($OSP$) for
$K^{} \Lambda$ channels from models
SL [1], and this work (A, B, and C).
The reduced $\chi^2$'s are given in the last row.
Model A is a simplified version of the SL model with $N1$ (spin 1/2)
and $N8$ (spin 5/2) resonances removed.
All the baryonic resonances have spin 1/2, except $N7$ and $L8$ (spin 3/2)
for which off-shell treatment is applied (models B and C). 
Model D is identical to model B,
with fixed values $X=-1/2$, $Z=0$, and $Y$ free.}
\protect\label{tab:ABC}
\begin{center}

\begin{tabular}{lcccrrrrr}
 &  &  & & &  & & & \\
{ Notation} &{ particle} & $(\ell)J^{\pi}$ &{ coupling} &
\multicolumn{1}{c}{SL} & \multicolumn{1}{c}{A} &
 \multicolumn{1}{c}{B} & \multicolumn{1}{c}{C} & \multicolumn{1}{c}{D}
\\
 &  &  &\multicolumn{1}{c} {{ and $OSP$}} &  &  &
 & &
\\\hline
 &  &  & & &  & & & \\
         & $\Lambda$ & ${1\over 2}^+$ & $g_{K\Lambda N}/ {\sqrt {4\pi}}$ &
 $-3.16 \pm 0.01$ &  $-3.16 \pm 0.01$ & $ -3.22 \pm 0.03$ & $ -3.22
\pm 0.01$
&  $-3.16 \pm 0.90$ \\[3pt]
         & ${\Sigma}$ & ${1\over 2}^+$ & $g_{K\Sigma N}/ {\sqrt {4\pi}}$ &
 $0.91 \pm 0.10$ &  $0.78 \pm 0.08$ &  $0.83 \pm 0.10$&  $0.86 \pm
0.02$
&  $0.87 \pm 0.06$ \\[3pt]
 &  &  & & &  & & & \\
$K^{*+}$ & $K^{*}(892)^+$ & $1^-$     & $G_V/ 4\pi$                      &
 $-0.05 \pm 0.01$&  $ -0.04 \pm 0.01$ &  $ 0.02 \pm 0.01$&  $ 0.02
\pm 0.01$
&  $0.02 \pm 0.01$ \\[3pt]
         &                &           & $G_T/ 4\pi$                      &
 $0.16 \pm 0.02$&  $ 0.18 \pm 0.02$ &  $ 0.18 \pm 0.01$&  $ 0.17
\pm 0.01$
&  $0.18 \pm 0.03$ \\[3pt]
$K1$     & $K1(1270)$     & $1^+$     & $G_{V1}/ 4\pi$                   &
 $-0.19 \pm 0.01$&  $-0.23 \pm 0.01$ &  $-0.15 \pm 0.01$ &  $-0.15
\pm 0.01$
&  $-0.17 \pm 0.01$ \\[3pt]
         &                &           & $G_{T1}/ 4\pi$                   &
 $-0.35 \pm 0.03$&  $-0.38 \pm 0.03$ &  $-0.38 \pm 0.04$&  $-0.39
\pm 0.03$
&  $-0.35 \pm 0.03$ \\[3pt]
 &  &  & & &  & & & \\
$N1$ & $ N(1440)$ & $(1){1\over 2}^+$ & $G_{N1}/ {\sqrt {4\pi}}$         &
 $-0.01 \pm 0.12$&  & &
&  {} \\[3pt]
$N7$ & $ N(1720)$ & $(1){3\over 2}^+$ & $G_{N7}^1/  4\pi$                &
 $-0.04 \pm 0.01$&  $ -0.04 \pm 0.01$ &  $ -0.04 \pm 0.01$& $-0.04
\pm 0.01$
&  $-0.03 \pm 0.01$ \\[3pt]
     &            &                   & $G_{N7}^2/  4\pi$                &
 $-0.14 \pm 0.04$&  $ -0.12 \pm 0.02$ & $-0.10 \pm 0.01$ &  $-0.10
\pm 0.01$
&  $-0.11 \pm 0.02$ \\[3pt]
     &            &                   &     X                            &
&  & $-1.03 \pm 0.21$ &  $-1.03 \pm 0.06$ &  $-0.5$ \\[3pt]
     &            &                   &     Y                            &
&  &  $8.25 \pm 0.28$&  $8.19 \pm 0.12$
&  $9.84 \pm 0.19$ \\[3pt]
     &            &                   &     Z                            &
&  & $ 0.003 \pm 0.014$&  $10^{-5} \pm 0.01$
&  $0.$ \\[3pt]
$N8$ & $ N(1675)$ & $(2){5\over 2}^-$ & $G_{N8}^a/  4\pi$                &
 $-0.63 \pm 0.10$&  & &
&  {} \\[3pt]
     &            &                   & $G_{N8}^b/  4\pi$                &
 $-0.05 \pm 0.56$&  & &
&  {} \\[3pt]
 &  &  & & &  & & & \\
$L1$&${\Lambda}(1405)$&$(0){1\over 2}^-$& $G_{L1}/ {\sqrt {4\pi}}$       &
 $-0.31 \pm 0.06$&  $ -0.29 \pm 0.05$ &  $ -0.28 \pm 0.02$&  $ -0.28
\pm 0.01$
&  $-0.29 \pm 0.05$ \\[3pt]
$L3$&${\Lambda}(1670)$&$(0){1\over 2}^-$& $G_{L3}/ {\sqrt {4\pi}}$       &
 $1.18 \pm 0.09$&  $ 1.15 \pm 0.13$ &  $1.26 \pm 0.02$&  $1.26 \pm
0.01$
&  $1.18 \pm 0.06$ \\[3pt]
$L5$&${\Lambda}(1810)$&$(1){1\over 2}^+$& $G_{L5}/ {\sqrt {4\pi}}$       &
 $-1.25 \pm 0.20$&  $-3.89 \pm 1.45$ &  $-1.78 \pm 0.05$&  $-1.78
\pm 0.02$
&  $-1.77 \pm 0.12$ \\[3pt]
$L8$&${\Lambda}(1890)$&$(1){3\over 2}^+$& $G_{L8}^1/4\pi$                &
&  & &  $0.002 \pm 0.045$
&  {} \\[3pt]
    &                 &                 & $G_{L8}^2/4\pi$                &
&  & &  $0.003 \pm 0.053$
&  {} \\[3pt]
     &            &                   &     X                            &
&  & & $-0.02 \pm 3.92$
&  {} \\[3pt]
     &            &                   &     Y                            &
&  & &  $0.23 \pm 9.20$
&  {} \\[3pt]
     &            &                   &     Z                            &
&  & &  $0.23 \pm 9.00$
&  {} \\[3pt]
 &  &  & & &  & & & \\
$S1$ & ${\Sigma}(1660)$ & $(1){1\over 2}^+$ & $G_{S1}/ {\sqrt {4\pi}}$   &
 $-4.96 \pm 0.19$ &  $-2.43 \pm 1.20$ &  $-5.37 \pm 0.05$&  $-5.36
\pm 0.02$
&  $-5.33 \pm 0.12$ \\[3pt]
%
\hline
$\chi^2$ &  &  & & \multicolumn{1}{c}{$1.73$}& \multicolumn{1}{c}{$1.84$}   &
\multicolumn{1}{c}{$1.66$} & \multicolumn{1}{c}{$1.69$}&
\multicolumn{1}{c}{$1.66$}\\
\end{tabular}
\end{center}

\end{table}
%
\bigskip
\begin{table}
\caption{Branching ratios ($BR\times 10^3$ in Eq.~[4.4]) 
for $K^- p \to \gamma \Lambda$, from the SL model
and the present work (models A, B, C, D).}
\protect\label{tab:BR}
\begin{center}
\begin{tabular}{lccccc}
 & & & & & \\
SL~[1]& A & B & C & D & experiment~[17] \\
 & & & & & \\ 
\hline
 & & & & &
\\[5pt]
 0.95 &  1.00 &  0.99 & 0.99 & 1.00 &0.86 $\pm$ 0.07 $\pm$ 0.09
\\ [5pt]
\end{tabular}
\end{center}
\end{table}
%
\begin{figure}
 
\caption{Differential cross section for the process 
$\gamma p \to K^+ \Lambda$:
excitation functions at
$\theta_{K}^{cm}$ = 27$^{\circ}$ (a), 90$^{\circ}$ (b) and
150$^{\circ}$ (c), and angular distribution at
$E_{\gamma}^{lab}$ = 1.0 GeV (d), 1.45 GeV (e), and 2.1 GeV (f). 
The curves are from models SL (dotted), A (dash-dotted), B (solid)
and C (dashed).
The SL model comes from Ref.~[1], and model A is a simplified version
of SL where the resonances $N1$ and $N8$ have been taken away
(see Table~I).
Model B is the same as model A, but with off-shell effects for the
only spin 3/2 resonance ($N7$) in the reaction mechanism. Model C
is the same as model B with an extra spin 3/2 (hyperonic) resonance 
($L8$), also with off-shell effects treatment. 
Data are from Refs. [10] (empty circles), and [11] (full circles).
}

\protect\label{fig:ds}
\end{figure}

\begin{figure}
 
\caption{Total cross section for the reaction 
$\gamma p \to K^+ \Lambda$
as a function of photon energy. Curves and data as in Fig.~1.}

\protect\label{fig:st}
\end{figure}

\begin{figure}
 
\caption{$\Lambda$-polarization asymmetry ($P$)
in $\gamma p \to K^+ \vec{\Lambda}$, polarized target 
asymmetry ($T$) in
$\gamma \vec{p} \to K^+ \Lambda$, and linearly 
polarized beam asymmetry 
($\Sigma$) in $\vec{\gamma} p \to K^+ \Lambda$: 
excitation functions at
$\theta_{K}^{cm}$ = 90$^{\circ}$ (a)-(c), and angular 
distributions at $E_{\gamma}^{lab}$ = 1.45 GeV (d)-(f)
and $E_{\gamma}^{lab}$ = 2.1 GeV (g)-(i). 
Curves are as in Fig. 1, and data from Refs. [12] ($P$), 
and [13] ($T$).}
  
\protect\label{fig:pol}
\end{figure}

\begin{figure}

\caption{Angular distributions for double polarization asymmetries 
($C_{x'}$, $C_{z'}$, $O_{x'}$, and $O_{z'}$)
in $\vec{\gamma} p \to K^+ \vec{\Lambda}$ 
at $E_{\gamma}^{lab}$ = 1.45 GeV (a)-(d)
and 2.1 GeV (e)-(h); curves as in Fig. 1.}

\protect\label{fig:coxz}
\end{figure}

\begin{figure}
 
\caption{Differential cross section $d\sigma_{UL}$ as a function of
momentum transfer ($Q^2$) for the reaction $e p \to e' K^+ \Lambda$,
for $W$ = 5.02 GeV$^2$, $t$ = $-0.15$ GeV$^2$, $\epsilon$ = 0.72.
Curves are as in Fig.~1, and the data from Ref.~[14].}

\protect\label{fig:ul}
\end{figure}

\begin{figure}
 
\caption{Same as Fig.~5, but for differential cross sections 
$d\sigma_{U}(t)$, $d\sigma_{L}(t)$, 
$d\sigma_{I}(t)$, and $d\sigma_{P}(t)$, see Eq.~(\ref{eq:sigel}). 
Letters T and L stand for transverse and longitudinal, respectively,
for $W$ = 5.02 GeV$^2$, $Q^2$ = 1 GeV$^2$,
and $\epsilon$ = 0.72. 
Curves are as in Fig.~1.}

\protect\label{fig:ulip}
\end{figure}

\begin{figure}
 
\caption{Same as Fig.~6, but for the 
longitudinal to transverse differential cross sections 
ratio $R(t)=d\sigma_L/d\sigma_U$. 
Curves are as in Fig.~1. 
}

\protect\label{fig:R}
\end{figure}

\begin{figure}
 
\caption{(a) Total cross section for the reaction 
$\gamma p \to K^+ \Lambda$
as a function of photon energy.
(b) Longitudinal to transverse differential cross sections 
ratio $R(t)=d\sigma_L/d\sigma_U$
for the reaction $e p \to e' K^+ \Lambda$.
The curves are from models B (solid) and D (dashed).
Model D has been obtained in the same conditions as model B,
except that the off-shell parameters X and Z were fixed at 
$-0.5$ and $0.0$, respectively (see Table I). 
}

\protect\label{fig:XZ}
\end{figure}
%

\begin{references}
%
\bibitem{SL}J. C. David, C. Fayard, G. H. Lamot,
 and B. Saghai, Phys. Rev. C {\bf 53}, 2613 (1996).
%
\bibitem{Ren} F. M. Renard and Y. Renard, 
Nucl. Phys. {\bf B25}, 490 (1971); 
Y. Renard, {\it ibid.\/} {\bf B40}, 499 (1972);
Y. Renard,
Th\`ese de Doctorat d'Etat \`es-Sciences Physiques,
Universit\'e des Sciences et Techniques du Languedoc, 
 1971 (in French).
%
\bibitem{Adel1}
R. A. Adelseck and L. E. Wright, Phys. Rev. C
{\bf 38}, 1965 (1988); R. A. Adelseck, Ph.D. Thesis, 
Ohio University, 1988.
%
\bibitem{Adel2} R. A. Adelseck, C. Bennhold, and L. E.
Wright, Phys. Rev. C {\bf 32}, 1681 (1985).
%
\bibitem{Ben89} M. Benmerrouche, R. M. Davidson, and Nimai C. Mukhopadhyay,
Phys. Rev. C {\bf 39}, 2339 (1989).
%
\bibitem{dav91} R. M. Davidson, Nimai C. Mukhopadhyay, and R. S. Wittman
Phys. Rev. D {\bf 43}, 71 (1991).
%
\bibitem{Ben95} M. Benmerrouche, Nimai C. Mukhopadhyay, and J. F. Zhang,
Phys. Rev. D {\bf 51}, 3237 (1995).
%
\bibitem{Ben92} M. Benmerrouche, PhD Thesis, Rensselaer Polytechnic
Institute (1992); 
Nimai C. Mukhopadhyay, J. F. Zhang, and M. Benmerrouche,
Proceedings of the {\it Forth CEBAF/INT
Workshop on N* Physics}, Seattle, WA, 9-13 Sept. 1996,
Editors: T.-S. H. Lee and W. Roberts, (World Scientific, 1997). 
%
\bibitem{BD} J. D. Bjorken and S. D. Drell, 
{\it Relativistic Quantum Mechanics} (McGraw-Hill, New York, 1964).
%
\bibitem{ST97} B. Saghai and F. Tabakin,  
Phys. Rev. C {\bf 55}, 917 (1997).
%
\bibitem{ade90} R. A. Adelseck and B. Saghai, Phys. Rev. C 
{\bf 42}, 108 (1990).
%
\bibitem{wil92} R. A. Williams, C. R. Ji, and S. R.
Cotanch, Phys. Rev. C {\bf 46}, 1617 (1992).
%
\bibitem{boc94} M. Bockhorst {\it et al.}, 
Z. Phys. {\bf C 63}, 37 (1994).
%
\bibitem{oldx} P. L. Donoho and R. L. Walker, Phys. Rev.  {\bf 112}, 
981 (1958);
B. D. McDaniel {\it et al.}, {\it ibid}  {\bf 115}, 1039 (1959);
H. M. Brody {\it et al.}, {\it ibid} {\bf 119}, 1710 (1960);
R. L. Anderson {\it et al.}, Phys. Rev. Lett. {\bf 9}, 131 (1962);
H. Thom {\it et al.}, {\it ibid} {\bf 11}, 434 (1963);
C. W. Peck, Phys. Rev.  {\bf 135}, 830 (1964);
R. L. Anderson {\it et al.}, {\it Proc. Int. Symp. on electron 
and photon 
interactions at high energies}, Hamburg, (1965);
S. Mori, Ph.D. Thesis, Cornell University (1966);
H. Thom, Phys. Rev. {\bf 151}, 1322 (1966);
D. E. Groom and J. H. Marshall, {\it ibid} {\bf 159}, 1213 (1967); 
R. Erbe {\it et al.}, {\it ibid} {\bf 188}, 2060 (1969);
A. Bleckmann {\it et al.}, Z. Phys. {\bf 239}, 1 (1970);
T. Fujii {\it et al.}, Phys. Rev. D {\bf 2}, 439 (1970);
D. Decamp {\it et al.}, Orsay Report, LAL 1236 (1970);
Th. Fourneron, Th\`ese de Doctorat d'Etat, Universit\'e
de Paris, Report LAL 1258, 1971, (in French);
H. Goeing {\it et al.}, Nucl. Phys. B{\bf 26}, 121 (1971);
P. Feller {\it et al.}, {\it ibid} B{\bf 39}, 413 (1972).
%
\bibitem{pol} B. Borgia {\it et al.}, Nuovo Cimento {\bf 32}, 218 
(1964); 
M. Grilli {\it et al.}, {\it ibid} {\bf 38}, 1467 (1965);
D. E. Groom and J. H. Marshall, Phys. Rev.  {\bf 159}, 1213 (1967);
R. Hass {\it et al.}, Nucl. Phys. B{\bf 137}, 261 (1978).
%
\bibitem{pot} K. H. Althoff {\it et al.}, Nucl. Phys. {\bf B137}, 
269 (1978).
%

\bibitem{dsige} C. N. Brown {\it et al.}, Phys. Rev. Lett. 
{\bf 28}, 1086, (1972); 
T. Azemoon {\it et al.}, Nucl. Phys. B {\bf 95}, 77 (1975); 
C. J. Bebek {\it et al.}, Phys. Rev. D {\bf 15}, 594
(1977); 
{\it ibid.\/} D {\bf 15}, 3082 (1977); 
P. Brauel {\it et al.}, Z. Phys. C {\bf 3}, 101 (1979).
%
\bibitem{BIGSKY} SAPHIR Collaboration (J. Barth {\it et al.})
Contributed paper to the {\it 6th Conference on the Intersections 
of Particle and Nuclear Physics}
(CIPANP 97), Big Sky, MT, 27 May - 2 June 1997 (nucl-th/9707025); 
D. Menze, {\it private communication}.
%
\bibitem{CROSY} C. R. Ji, and S. R. Cotanch, 
Phys. Rev. C {\bf 38}, 2691 (1988).
%
\bibitem{whi89} D. A. Whitehouse {\it et al.}, 
Phys. Rev. Lett. {\bf 63}, 1352 (1989).
%
\bibitem{st96} B.~Saghai and F.~Tabakin,  
Phys. Rev. C {\bf 53}, 66 (1996).
%
\bibitem{nat80} L. M. Nath and B. Bhattacharyya, 
Z. Phys. C {\bf 5}, 9 (1980).
%
\bibitem{lu95} D. Lu, R. H. Landau, and S. C. Phatak,
Phys. Rev. C {\bf 52}, 1662 (1995).
%
\bibitem{kai97} N. Kaiser, T. Waas, and W. Weise,
Nucl. Phys. A {\bf 612}, 297 (1997).
%
\bibitem{CHPT} S. Steininger and U.-G. Mei{\ss}ner,
Phys. Lett. {\bf B391}, 446 (1997).
%
\bibitem{ZPL} Zhenping Li, Phys. Rev. C {\bf 52}, 1648 (1995);
Zhenping Li, Hongxing Ye, and Minghui Lu, {\it ibid.}
C {\bf 56}, 1099 (1997).
%
\end{references}
\end{document}